\documentclass[printer]{aa} 
\usepackage{threeparttable}
\usepackage{lscape}
\usepackage{multirow}
\usepackage[normalem]{ ulem }
\usepackage{soul}
\usepackage{colortbl}
\usepackage{grffile, graphicx}
\usepackage[T1]{fontenc}
\usepackage{lmodern}
\usepackage{txfonts}

\newcommand{\rot}{\rotatebox{90}}
\newcommand{\Teff}{\ensuremath{\mathrm{T_{eff}}}}
\newcommand{\MJup}{\ensuremath{\mathrm{M_{Jup}}}}
\newcommand{\RJup}{\ensuremath{\mathrm{R_{Jup}}}}

\newcommand{\Msun}{\ensuremath{\mathrm{M_{\odot}}}}
\newcommand{\mic}{\ensuremath{\mathrm{\mu m}}}
\newcommand{\Ha}{\ensuremath{\mathrm{H_{\alpha}}}}
\newcommand{\Pab}{\ensuremath{\mathrm{Pa_{\beta}}}}
\newcommand{\Hb}{\ensuremath{\mathrm{H_{\beta}}}}
\newcommand{\Hd}{\ensuremath{\mathrm{H_{\delta}}}}
\newcommand{\Hg}{\ensuremath{\mathrm{H_{\gamma}}}}
\newcommand{\Lum}{\ensuremath{\mathrm{log(L/L_{\odot})}}}
\newcommand{\Lumacc}{\ensuremath{\mathrm{log(L_{acc}/L_{\odot})}}}
\DeclareUnicodeCharacter{2212}{-}

\begin{document} 
   \title{A new take on the low-mass brown dwarf companions on wide-orbits in Upper-Scorpius\thanks{Based on observations collected at the European Organisation for Astronomical Research in the Southern Hemisphere under ESO program 093.C-0769.}}
\titlerunning{X-Shooter view of wide orbit companion in Upper-Sco}

   \author{S. Petrus\inst{1}, M. Bonnefoy\inst{1}, G. Chauvin\inst{2,1}, C. Babusiaux\inst{1}, P. Delorme\inst{1}, A.-M. Lagrange\inst{1}, N. Florent\inst{1}, A. Bayo\inst{3,4}, M. Janson\inst{5}, B. Biller\inst{6}, E. Manjavacas\inst{7}, G.-D. Marleau\inst{8}, T. Kopytova\inst{9,10}}
   \authorrunning{Petrus et al.}

   \institute{\inst{1} Univ. Grenoble Alpes, CNRS, IPAG, 38000 Grenoble, France \\
   \inst{2} Unidad Mixta Internacional Franco-Chilena de Astronom\'{i}a, CNRS/INSU UMI 3386 and Departamento de Astronom\'{i}a, Universidad de Chile, Casilla 36-D, Santiago, Chile\\
   \inst{3} Instituto de F\'isica y Astronom\'ia, Universidad de Valpara\'iso, Chile\\
   \inst{4} N\'ucleo Milenio de Formaci\'on Planetaria (NPF), Chile\\
   \inst{5} Stockholm University Department of Astronomy AlbaNova University Center 10691 Stockholm Sweden\\
   \inst{6} Institute for Astronomy The University of Edinburgh Royal Observatory Blackford Hill Edinburgh EH9 3HJ U.K.\\
   \inst{7} W. M. Keck Observatory 65-1120 Mamalahoa Hwy. Kamuela, HI 96743 USA\\
   \inst{8} Institut f\"ur Astronomie und Astrophysik, Eberhard Karls Universit\"at T\"ubingen, Auf der Morgenstelle 10, 72076 T\"ubingen, Germany\\
   \inst{9} School of Earth \& Space Exploration, Arizona State University, Tempe AZ 85287, USA\\
   \inst{10} Ural Federal University, Yekaterinburg 620002, Russia}

   \date{Received 19/04/2019; accepted 27/09/2019}

  \abstract
   {The Upper-Scorpius association (5-11 Myr) contains a unique population of low-mass ($\rm{M}\leq30$\,\MJup) brown-dwarfs either free-floating, forming wide pairs, or on wide-orbits to solar-type and massive stars. The detailed relative characterization of their physical properties (mass, radius, temperature, composition, ongoing accretion) offers the opportunity to potentially explore their origin and their mechanisms of formation.}
    {In this study, we aim at characterizing the chemical and physical properties of three young, late-M brown-dwarfs claimed to be companions of the Upper-Scorpius stars USco\,161031.9-16191305, HIP\,77900, and HIP\,78530 using medium resolution spectroscopy at UV ($0.30-0.56$\,\mic; $R_{\lambda}\sim3300$), optical ($0.55-1.02$\,\mic; $R_{\lambda}\sim5400$), and near-infrared ($1.02-2.48$\,\mic; $R_{\lambda}\sim4300$) wavelengths. The spectra of six free-floating analogues from the same association are analyzed for comparison and to explore the potential physical differences between these substellar objects found in different configurations. We also aim at looking and analyzing hydrogen emission lines at UV and optical wavelengths to investigate the presence of ongoing accretion processes.}
   {The X-Shooter spectrograph at VLT was used to obtain the spectra of the nine young brown dwarfs over the  $0.3-2.5$\,\mic\,range simultaneously. Performing a forward modelling of the observed spectra with the \texttt{ForMoSA} code, we infer the \Teff, log(g), and radius of our objects. The code compares here the \texttt{BT-SETTL15}\, models to the observed spectra using the Nested Sampling Bayesian inference method. Mass is determined in using evolutionary models and a new analysis of the physical association is proposed in using the \textit{Gaia}-DR2 astrometry.}
   {The \Teff\ and log(g) determined for our companions are compatible with those found for free floating analogues of the Upper-Scorpius association and with evolutionary-model predictions at the age of the association. However the final accuracy on the \Teff\,estimates is strongly limited by non-reproducibilities of the \texttt{BT-SETTL15} models in the range of \Teff\,corresponding to the M8-M9 spectral types.  We identified \Ha, \Hb, \Hg, and Ca II H \& K emission lines in the spectrum of several objects. We attribute these lines to chromospheric activity except for the free-floating object USco\,1608-2315 for which they are indicative of active accretion ($\mathrm{\dot{M} \leq 10^{-10.76} M_{\odot} /year}$). We confirm the $\times$ 4 over-luminosity of USco\,161031.9-16191305\,B down to 0.3 $\mu$m. It could be explained in part by the object activity and if the companion is an unresolved multiple system.}
   {}

   \maketitle

\section{Introduction}
The first brown dwarfs (BDs) were contemporaneously discovered with the first exoplanets at the end of the last millennium \citep{1995Natur.378..355M, 1995Natur.378..463N,  1995Natur.377..129R}. Since then, thousands of BDs have been detected and studied in isolation \citep{2005ApJ...623.1115C, 2005ARAeA..43..195K} in the field or as wide- or short-period companions to nearby stars \citep{2011AeA...525A..95S, 2012IAUS..282..105A}. Bridging the gap between planets and stars, BDs are too light by definition to burn Hydrogen but massive enough to burn  Deuterium \citep{1997ApJ...491..856B}. Despite two decades of intensive study, many fundamental questions remain unanswered regarding their  formation and evolution processes, their physical and atmosphere properties, and their connection to stars and planets.  Multiple stellar-like formation pathways have been proposed for these objects: i/ turbulent fragmentation of molecular clouds \citep{2004ApJ...617..559P}, ii/ premature ejection of protostellar embryos \citep{2002MNRAS.332L..65B}, iii/ photo-erosion of prestellar cores \citep{1996AJ....111.2349H}, and iv/ disk-instability \citep[e.g.,][]{2009MNRAS.392..413S}. Studies of young star-forming regions are currently on-going to identify differences in binary statistics linked to these different processes \citep[e.g.]{2017AeA...605A..11M,2015ApJ...800...72T}. Alternatively, we know from observations and theory that planetary formation mechanisms like core accretion are probably forming very massive giant planets and populating the mass distribution of substellar companion up to $\mathrm{M \leq 35\,M_{Jup}}$  \citep[e.g., ][]{2009AeA...501.1139M, 2012AeA...541A..97M}. There is therefore little doubts that both stellar and planetary-like formation mechanisms overlap in mass distribution and an interesting question is to investigate whether stellar and planetary mechanisms might lead to different atmospheric properties that could be traced through observations. Core accretion might indeed lead to an overabundance of heavy elements in the atmosphere. Our ability to identify such chemical imprint is very challenging, and beyond the observing limitations, directly connected to our understanding and modelling of the physics of brown dwarf and exoplanet atmospheres. Confronting the latest predictions of substellar atmosphere models with high-quality optical and infrared spectra of young brown dwarfs is a key step toward this goal.

Evolutionary models predict how BDs contract and cool down with time, and how their fundamental parameters such as effective temperature, surface gravity, radius, and luminosity evolve \citep{2003AeA...402..701B}. The contraction leads to a decrease of the radius and an increase of the surface gravity. This evolution impacts the bolometric luminosity and the spectrum with a modification of the pseudo-continuum and the appearance of atomic and molecular absorption lines at different evolutionary stages. The spectral morphology has been used for years to extend the old stellar spectral classification into the substellar one, from M-dwarfs to L, T and Y-dwarfs reaching effective temperatures as cold as  \citep[\Teff\,$\leq$ 450K;][]{2012ApJ...753..156K, 2014ApJ...797....3K}. The fine characterization of the BD atmospheres therefore improves our global understanding of their physical properties, and also their formation and their evolution.

There are currently two approaches for the spectral characterization of BDs. The first one is empirical and based on the comparison with libraries of known young and old BDs \citep[e.g.,][]{2013ApJ...772...79A, 2014AeA...562A.127B, 2017MNRAS.465..760B}, and is tightly connected and limited by the size and the diversity of these libraries. A complementary alternative is the comparison of the observed spectra to the recent models of substellar atmospheres \citep[e.g.,][]{2001ApJ...556..357A, 2006AeA...455..325H,2011ApJ...735L..39B, 2011ApJ...737...34M, 2012ApJ...756..172M, 2017ApJ...850...46T, 2018ApJ...854..172C} of BD and giant planets. This offers the advantage to derive the physical parameters for a given model independently of other observations and to test the influence of new physical ingredients (new atomic/molecular line opacities, presence and properties of clouds, non-equilibrium chemistry, thermo-chemical instability...). Despite this, it suffers from our limited observational knowledge and constraints to disentangle their relative importance leading at the end to significant degeneracies and systematics in the fundamental parameters derived when using different families of atmosphere models or simply different set of physical parameters. Both approaches remain therefore very complementary today.  

The detection of forming companions is a new observational window on the initial conditions of planetary systems \citep[formation zone, timescales, and modes; accretion physics; e.g.,][]{2017AeA...608A..72M}. \Ha\,(656.3 nm) and \Pab\,(1282.2 nm) lines have been detected in the spectra of 8 companions with masses below 30 \MJup\,and ages in the 1-10 Myr range thus far \citep[TWA 5 B, GQ Lup B, CT Cha B, USco CTIO 108 B, DH Tau B, GSC 06214-00210 B, SR 12 C, PDS 70 B][]{2000AeA...360L..39N, 2007AeA...463..309S,2008AeA...491..311S,2008ApJ...673L.185B, 2014AeA...562A.127B, 2011ApJ...743..148B, 2018MNRAS.475.2994S, 2018ApJ...863L...8W, 2014ApJ...783L..17Z, 2017ApJ...836..223W}. These lines are known tracers of active accretion and of sub-stellar chromospheric activity. Sub-millimeter observations of these objects have failed to reveal excess emission from the expected mass reservoir (circumplanetary disk) surrounding these objects thus far \citep{2010AJ....139..626D, 2015ApJ...805L..17B, 2017ApJ...835...17M, 2017AJ....154...26W, 2017AJ....154..234W, 2017ApJ...836..223W} and only one companion GSC 06214-00210 B shows clear excess emission at near-infrared wavelengths \citep{2011ApJ...743..148B}. This might indicate these companions bear very compact and optically thick disks \citep{2017ApJ...836..223W}.

 The Upper Scorpius subgroup (hereafter Upper Sco) in the Scorpius-Centaurus OB association contains one of the nearest \citep[$d = 146\pm3$\,pc;][]{1999MNRAS.310..585D, 1999AJ....117..354D, 2018MNRAS.477L..50G} and richest population of young stars, and substellar objects \citep[e.g.][]{2000AJ....120..479A, 2007MNRAS.374..372L, 2018MNRAS.473.2020L, 2018AJ....156...76L} down to the planetary-mass range. The extinction is low in this region \citep[$A_V\leq2$ mag;][]{1994AJ....107..692W, 2019arXiv190204116L}. At an estimated age of 5 to 11 Myr \citep{2012ApJ...746..154P, 2019ApJ...872..161D}, stars harbor primordial, transitional, and debris disks \citep{2012ApJ...758...31L, 2018AJ....156...75E} suggesting planet/BD formation within disks at different completion levels. Upper Sco also contains a large sample of low-mass BDs ($\rm{M}<30$\,\MJup) and planetary mass companions identified with deep-imaging and high-contrast imaging techniques. These companions are found over a wide range of projected separations \citep[$\sim$300-3400 au;][]{2008ApJ...673L.185B, 2008ApJ...689L.153L, 2010ApJ...719..497L, 2011ApJ...730...42L, 2011ApJ...726..113I, 2013ApJ...773...63A} from M7 to B6-type stars. Some of the free-floating low-mass brown-dwarfs and companions harbor disks \citep[e.g., ]{2011ApJ...743..148B, 2013MNRAS.429..903D} and are actively accreting \citep{2009ApJ...696.1589H, 2011ApJ...743..148B, 2018MNRAS.473.2020L}. This unique population of low mass BDs and planetary-mass objects with  various configurations (as companions, binaries, free-floating) and likely diverse origins,  represents a unique testbed for planet and BD formation models. 

In this paper, we present a new study of three young brown dwarfs companions to the stars HIP\,78530, HIP\,77900, and USco\,161031.9-16191305 (hereafter USco\,1610-1913), members of the Upper Scorpius association. We  obtain VLT/X-Shooter 0.3-2.5 \mic\, spectra of these objects and use them to characterize their physical properties. The data notably extend previous analysis of these objects to the optical at medium resolving powers, thus enabling for an investigation of  emission lines related to accretion and for testing the atmospheric models of young BDs. We provided an up-to-date description of our targets in Section \ref{sec:Targets}. The observations and the data reduction are detailed in Section \ref{Sec:Observations}. In Section \ref{Sec:Physical properties}, we present our results using both empirical and synthetic model approaches. Our \texttt{ForMoSA} forward modelling tool is used to explore the different atmospheric models and determine the most probable physical properties of these three companions. In Section \ref{sec:Emission line properties}, we focus our study on the emission line properties observed for these three companions. In Section \ref{Sec:Discussion}, we finally summarize and discuss our results in the context of previous work, update from the \textit{Gaia} Data Release 2, and perspective of future studies. 

\section{Target description}
\label{sec:Targets}
    
 \begin{table*}[t]
\caption{Observing log}
\label{tab:obs}
\renewcommand{\arraystretch}{1.3}
\begin{center}
\small
\begin{tabular}{l|llllllllll}
\hline
\hline
Target 	&	Date	&	UT Start-Time  &  $\mathrm{DIT}$ & $\mathrm{NDIT}$  & $\mathrm{NEXP}$  & $<$Seeing$>$ &  Airmass	&	Notes 	\\
		& (yyyy-mm-dd)	&			(hh:mm)	&	(s)						&							  &						    	 &      (")                &	             &                \\
\hline
USco\,161031.9-16191305\,B		&	2014-04-02	&	08:45	&	190/190/200	&	1/1/1	&	8/8/8	&	0.75	&	1.02	&	\\
HIP\,77900\,B		&		2014-04-12	& 05:45 &	190/190/197	&	1/1/1	    &	14/14/14	&	1.14	&	1.03	&	\\
					    &		2014-04-12		&	 08:38  & 190/190/197 &	1/1/1	    &	14/14/14	&	1.01	&	1.10	&	\\
HIP\,78530\,B		&		2014-06-09	& 04:08 	&	190/190/197	&	1/1/1	    &	14/14/14	&	0.50	&	1.03	&	\\
	&		2014-06-09	& 05:14 	&	190/190/197	&	1/1/1	    &	14/14/14	&	0.85	&	1.15	&	\\
\hline
USco\,J160723.82-221102.0	&	2014-04-22	&	09:01	&	190/190/197	&	1/1/1	& 8/8/8 & 		0.73	&	1.20	&	\\
	&	2014-06-14	&	23:50	&	190/190/197	&	1/1/1	& 14/14/14 & 	0.52	&	1.33	&	\\
USco\,J160606.29-233513.3	&	2014-06-25	&	04:13	&	190/190/197	&	1/1/1	&	14/14/14	&	0.86	&	1.14	&	\\
	&	2014-06-30	&	03:09	&	190/190/197	&	1/1/1	&	14/14/14	&	0.69	&	1.06	&	\\
	&	2014-07-04	&	01:05	&	190/190/197	&	1/1/1	&	16/16/16	&	0.65	&	1.02	&	\\
	&	2014-07-29	&	00:44	&	190/190/197	&	1/1/1	&	16/16/16	&	0.67	&	1.03	&	\\
USco\,J161047.13-223949.4 &	2014-06-30	&	04:21	&	190/190/197	&	1/1/1	&	14/14/14	&	0.88	&	1.20	&  [1]\\
 &	2014-07-01	&	03:26	&	190/190/197	&	1/1/1	&	6/6/6	&	1.03	&	1.06	& \\
 &	2014-07-01	&	04:04	&	190/190/197	&	1/1/1	&	14/14/14	&	1.14	&	1.16	& \\
USco\,J160737.99-224247.0	&	2014-07-04	&	05:07	&	190/190/197	&	1/1/1	&	14/14/14	&	1.38	&	1.51	& \\
	&	2014-07-29	&	02:07	&	190/190/197	&	1/1/1	&	14/14/14	&	0.61	&	1.16	& \\
	&	2014-07-29	&	03:08	&	190/190/197	&	1/1/1	&	14/14/14	&	0.54	&	1.39	& \\
	&	2014-08-02	&	01:49	&	190/190/197	&	1/1/1	&	16/16/16	&	1.33	&	1.16	& \\
	&	2014-08-02	&	03:09	&	190/190/197	&	1/1/1	&	16/16/16	&	1.12	&	1.51	& \\
USco\,J160818.43-223225.0	&	2014-07-02	&	03:16	&	190/190/197	&	1/1/1	&	14/14/14	&	1.11	&	1.07	& \\
	&	2014-07-02	&	04:26	&	190/190/197	&	1/1/1	&	14/14/14	&	0.80	&	1.25	& \\
	&	2014-07-03	&	03:24	&	190/190/197	&	1/1/1	&	14/14/14	&	1.03	&	1.10	& \\
	&	2014-07-04	&	03:49	&	190/190/197	&	1/1/1	&	14/14/14	&	1.14	&	1.16	& \\
USco\,J160828.47-231510.4	&	2014-06-15	&	01:00	&	190/190/197	&	1/1/1	&	14/14/14	&	0.95	&	1.11	& \\
	&	2014-06-20	&	05:06	&	190/190/197	&	1/1/1	&	14/14/14	&	1.20	&	1.23	& \\
\hline
\hline
\end{tabular}
\end{center}
\tablefoot{The seeing is measured at 0.5\,\mic\,and given for the visible arm. The DIT (Detector Integration Time) values refer to the individual exposure time per frame in the UVB, VIS, and NIR arms respectively. NDIT are the number of individual frames per exposure, and $N_{EXP}$  the number of exposures  in the UVB, VIS, and NIR arms. [1] no STD observed.}
\end{table*}

	 \begin{table*}[t]
\caption{Description of the properties of the three systems HIP\,78530, HIP\,77900, and USco\,1610-1913}
\label{tab:comp descr}
\small
\renewcommand{\arraystretch}{1.3}
\begin{threeparttable}
  \begin{tabular}{c||cc|cc|cccc|c}
\cline{4-9}
   \multicolumn{3}{c}{}  &  \multicolumn{2}{|c}{Primary}  &  \multicolumn{4}{|c|}{Companion} \\
\hline
 Source  &  d\tnote{\textit{c}}\, (pc)  &   $\mathrm{A_{v}}$\tnote{\textit{d}}\, (mag)   &  \Teff\, (K)  &  SpT  &  \Teff\, (K)  &  SpT   & Mass (M$_{Jup}$) & Separation\, (AU)  &   Ref  \\
\hline	
\hline	
USco\,1610-1913\,B  & 143.9$\pm$8.0 & 0.13  & 4140$\pm$150   & K7       & 2400$\pm$150 & M9$\pm$0.5 & 20$\pm$5  & 779$\pm$9 & \textit{a}  \\
HIP\,77900\,B      & 150.8$\pm$3.0 & 0.07  & 13700$\pm$1500 & B6$\pm$1 & 2400$\pm$150 & M9$\pm$0.5 & 20$\pm$7  &  3200$\pm$300 & \textit{a}  \\
HIP\,78530\,B      & 137.2$\pm$1.5 & 0.075 & $\simeq$10500  & B9V      & 2700$\pm$100 & M7$\pm$0.5 & 23$\pm$2  & $623\pm$8    & \textit{b}  \\
\hline
\end{tabular}
\begin{tablenotes}
\item References: \textit{a} \cite{2013ApJ...773...63A}; \textit{b} \cite{2015ApJ...802...61L}; \textit{c} \cite{2018MNRAS.477L..50G}; \textit{d} \cite{2019arXiv190204116L}.
\end{tablenotes}
\end{threeparttable}
\end{table*}

The three companions have close physical properties and spectral type (M8-M9), yet they were selected because they come with different configurations (mass ratio with the host star, projected separation. See Table \ref{tab:comp descr}):

\begin{itemize}
\item HIP\,78530\,B was identified by \cite{2005AeA...430..137K} at a separation of $4.536\,\pm\,0.006\,\!''$ from the $\sim$2.5\,\Msun\, B9V star \citep{1988mcts.book.....H} HIP\,78530\,A. \cite{2011ApJ...730...42L} confirmed the companion is co-moving with the primary star. The projected separation between the two objects now corresponds to $623\,\pm\,8$\,au using the \textit{Gaia}-DR2 distance ($137.2\,\pm\,1.5$pc; \textit{Gaia}\, Collaboration, 2018). The \texttt{Banyan $\Sigma$} tool \citep{2018ApJ...856...23G} and the DR2 astrometry confirms the star is a high-probability member (99.9\%) of the Upper-Scorpius association. \cite{2011ApJ...730...42L} has provided a medium-resolution ($R_\lambda\sim5300$ to 6000) spectrum of the companion covering the $1.15-2.40$\,\mic\, wavelength range. \cite{2015ApJ...802...61L} also presented a lower-resolution spectrum  ($R_\lambda\sim1350$) but extending down to 1 \mic. The spectra confirm  the companion is a young M7$\,\pm\,$0.5  dwarf. \cite{2015ApJ...802...61L} estimate a \Teff\, of $2700\,\pm\,100$\,K and a luminosity of \Lum$\,=\,-2.53\,\pm\,0.09$\, relying on the \textit{Hipparcos} distance available at that time. They find a mass of $23\,\pm\,2$\,\MJup\, assuming an age of 10 Myr for Upper-Scorpius. The primary star has no noticeable  excess emission \citep{2009ApJ...705.1646C, 2012ApJ...758...31L, 2017MNRAS.471..770M} and low extinction \citep[$A_{v}\,=\,0.48$][]{1980AeAS...42..251N}. 
 
\item HIP\,77900\,B was identified in UKIDSS and Pan-STARRS 1 images by \cite{2013ApJ...773...63A} from its red colors. It is located at a projected separation of $21.8\,\!''$ from the B6 star \citep{1967ApJ...147.1003G} HIP\,77900, a high probability member of Upper-Sco (97.6\% membership probability according to \texttt{Banyan $\Sigma$}). Unlike the case of  HIP\,78530\,B, the authors did not re-observe the target to check whether it shares the primary star's proper motion.  However, they obtained a low-resolution ($R_\lambda\sim100$)  spectrum of the source covering the $0.8-2.5\,$\mic\,range whose features are indicative of a young M9 ($\,\pm\,0.5$) object from Upper-Sco. Therefore, they argued HIP\,77900\,B to be bound to the star.  We re-discuss the physical association of the two objects in Section \ref{subsec:physasso}.  

\item USco\,1610-1913\,B was identified by \cite{2008ApJ...686L.111K} and  confirmed to be co-moving with the K7 star GSC 06209-00691  \citep{2008ApJ...686L.111K, 2009ApJ...703.1511K, 2014ApJ...781...20K}. The companion was last found at  $5.837\,\pm\,0.006\,\!''$ from the star, now corresponding to a separation of $779\,\pm\,9$\,au at the \textit{Gaia}-DR2 distance of the primary.  \cite{2013ApJ...773...63A} presented a $0.8-2.5$\,\mic\, low-resolution ($R_\lambda\sim100$) spectrum of the object confirming its youth and substellar nature. They estimated a M$9\,\pm\,0.5$ spectral type and found the companion to be four times overluminous with respect to HIP\,77900\,B which shared the same spectral properties at this spectral resolution. The spectral type was confirmed by \cite{2015ApJ...802...61L} from a medium-resolution ($R_\lambda\sim5300$ to 6000) near-infrared ($1.15-2.40\,$\mic) spectrum of the companion. They find \Teff\,in the range $2300-2700$\,K for the object using \texttt{DRIFT-PHOENIX} and \texttt{BT-SETTL} atmospheric models with a large spread in \Teff\,values depending on the wavelength interval considered for the fit. Their analysis also confirms the over-luminosity of the object for the inferred temperature. However, the star had no measured parallax at that time so that the average distance of Upper-Sco from \cite{1999AJ....117..354D} was assumed. A second companion (hereafter USco\,1610-1913\,Ab) was discovered at a projected separation of $19.4\,\pm\,0.3$\,au \citep{2008ApJ...679..762K} from USco\,1610-1913\,A. USco\,1610-1913\,Ab has a mass of $103\,\pm\,24$\,\MJup\,using the contrast reported in  \cite{2008ApJ...679..762K}, the new \textit{Gaia}-DR2 distance of USco\,1610-1913\,A (assuming it is unbiased; see section \ref{subsec:physasso}), and the  \cite{2015AeA...577A..42B} tracks at an age of 5 to 11 Myr.

The six young and isolated free-floating objects (see Table \ref{tab:obs})  were selected from the sample of \cite{2008MNRAS.383.1385L} late-M/early-L Upper-Sco brown dwarfs who also report low-resolution ($R_\lambda\sim1700$) spectra over $1.15-2.50$\,\mic\, of the sources. Low-resolution ($R_\lambda\sim1350$) multi-epoch red-optical ($0.57-0.88$\,\mic) spectra of these objects have also been collected by \cite{2011AeA...527A..24L}. The spectra of USco\,J1610-2239 and USco\,J1607-2211 exhibit a \Ha\,line. No emission line  was detected at that time in the spectra of the remaining objects. \cite{2013MNRAS.429..903D} and \cite{2012ApJ...758...31L} do not find noticeable infrared excess (up to 22 \mic) for USco\,J1610-2239,  USco\,J1607-2211,  USco\,J1607-2242. This is not the case of USco\,J1606-2335 and USco\,1608-2315. The former is found to have Spitzer [4.5] and WISE W2 photometry indicative of a disk excess \citep{2012ApJ...758...31L}.  The latter is found to have a W1-W2 color suggestive of an excess \citep{2013MNRAS.429..903D} but that excess is not confirmed by \cite{2012ApJ...758...31L} using similar data. USco\,J1610-2239 and USco\,J1607-2211 have \textit{Gaia}-DR2 distances of $128.5^{+15.4}_{-12.4}$\,pc and  $119.3^{+20.8}_{-15.4}$\,pc. Using \texttt{Banyan $\Sigma$} \citep{2018ApJ...856...23G}, we find that the two objects have a 99.6\% and 94.9\% chance respectively of belonging to Upper-Scorpius based on kinematics.   USco\,J1608-2315 also has a \textit{Gaia}-DR2 parallax value ($\pi\,=\,4.5073\,\pm\,1.1874$\,mas). The object is found at a larger distance than the typical Upper-Sco members and therefore \texttt{Banyan $\Sigma$} estimates a lower probability of membership to the association (59.7\%). The large error of DR2 for this object leads us to doubt this probability. Noting that the target has spectral features clearly indicative of a membership to Upper-Sco \citep[e.g.,][]{2008MNRAS.383.1385L, 2014AeA...562A.127B}. To conclude, adaptive-optics imaging and sparse aperture masking observations \citep{2011ApJ...730...39B, 2012ApJ...757..141K} of the sources exclude companions with similar masses down to $\sim$10 au. 
 \end{itemize}
 
 The optical and near-infrared spectra of the free-floating sources presented here have already been briefly introduced in \cite{2018MNRAS.473.2020L}. They  have increased spectral resolution or are extended at shorter wavelengths compared to previous spectroscopic data, and thus allow us to study the long-term variability of  the \Ha\, line. We therefore re-analyse them to use them as empirical templates for the characterization of the three companions HIP\,78530\,B, HIP\,77900\,B, and USco\,1610-1913\,B. 

\section{Observations and data reduction}
\label{Sec:Observations}
We used the X-Shooter seeing-limited  medium resolution spectrograph  mounted at UT2 Cassegrain focus \citep{2011AeA...536A.105V}. The wide wavelength coverage of the instrument (300-2480\,nm) is ideally suited for the characterization of accreting brown dwarfs with emission line series.  We chose the 1.6", 1.5", and 1.2"-wide slits for the  UVB, VIS, and NIR arms, corresponding to resolving powers $R_{\lambda}\,=\,\lambda/\Delta \lambda\,=\,$3300, 5400, and 4300, respectively. This set up was adopted for all our targets.  The observing log is reported in Table \ref{tab:obs}. The slits were oriented perpendicular to the companion's position angles in order to mitigate the flux contamination of the host stars. Each target was observed following a ABBA strategy to evaluate and remove the sky emission at the data processing step. Spectro-photometric standard stars were observed as part of the observatory calibration plan. They are not reported in the log as they were not used for our reduction and analysis.

We used the ESO \texttt{reflex} data reduction environment  \citep{2013AeA...559A..96F} to run the X-Shooter pipeline version 2.9.3 on the raw data \citep{2010SPIE.7737E..28M}. The pipeline produces two-dimensional, curvature-corrected, and flux-calibrated spectra  for each target and epoch of observation (trace). The spectra were extracted from the traces using a custom IDL script. The flux in each wavelength channel at the position of the source was averaged within  720 mas aperture in the UVB and VIS arms, and a 1120 mas aperture in the NIR arm. The script computed the noise at the position of the source into each spectral channel following the procedure described in \cite{2017AeA...602A..82D}. The residual non-linear pixels in the spectra were removed using the kappa-sigma clipping method. The telluric correction were evaluated and removed using the \texttt{molecfit} package \citep{2015AeA...576A..77S, 2015AeA...576A..78K}. The spectra at each epoch were corrected from the barycentric velocity and re-normalized using the epochs when the  sky transmission was photometric as an anchor point.  Our flux-calibration was checked computing the 2MASS or MKO synthetic photometry from the spectra and comparing the values to published ones \citep{2008MNRAS.383.1385L}. 

\section{Physical properties}
\label{Sec:Physical properties}
	\subsection{Empirical analysis}
	\label{Sec:Empirical_analysis}

\begin{figure}[t]
  \centering
  \includegraphics[width=\columnwidth, angle=0]{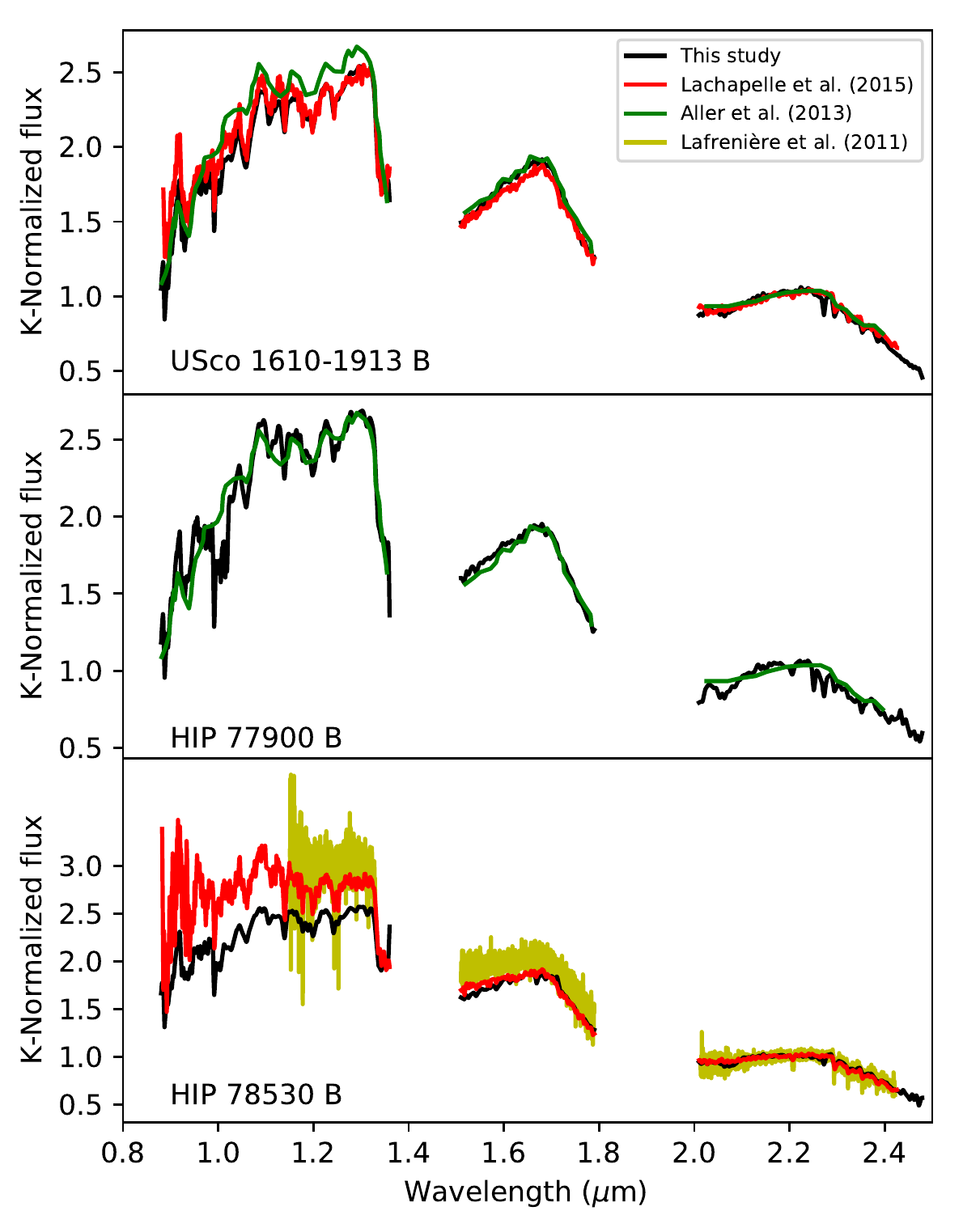}
  \caption{Comparison between our spectra (\textit{black} from X-Shooter), the ones from \cite{2015ApJ...802...61L} in \textit{red} (R = 900) and \cite{2013ApJ...773...63A} in \textit{green} (digitized) and the one from \cite{2011ApJ...730...42L} in  \textit{yellow} (R = 5300-6000).}
  \label{Fig:CompLC}
\end{figure}

	\cite{2015ApJ...802...61L} derived the spectral types of USco\,1610-1913\,B and HIP \,78530\,B using near-infrared spectra of these companions. We noticed significant differences between them and the X-Shooter spectrum of HIP\,78530\,B presented here (see Figure \ref{Fig:CompLC}). We noticed the same difference for the same target in comparing with the spectrum from \cite{2011ApJ...730...42L}. The X-Shooter spectrum of USco\,1610-1913\,B is consistent with the low-resolution spectrum of \cite{2013ApJ...773...63A}. We therefore conclude that the differences in the J-band spectrum of HIP\,78530\,B may stems from flux losses affecting long-slit observation with AO-fed spectrographs.

	    \subsubsection{Line identification}
 \begin{figure*}[t]
  \centering
  \begin{tabular}{cc}
  \includegraphics[width=8cm, angle=0]{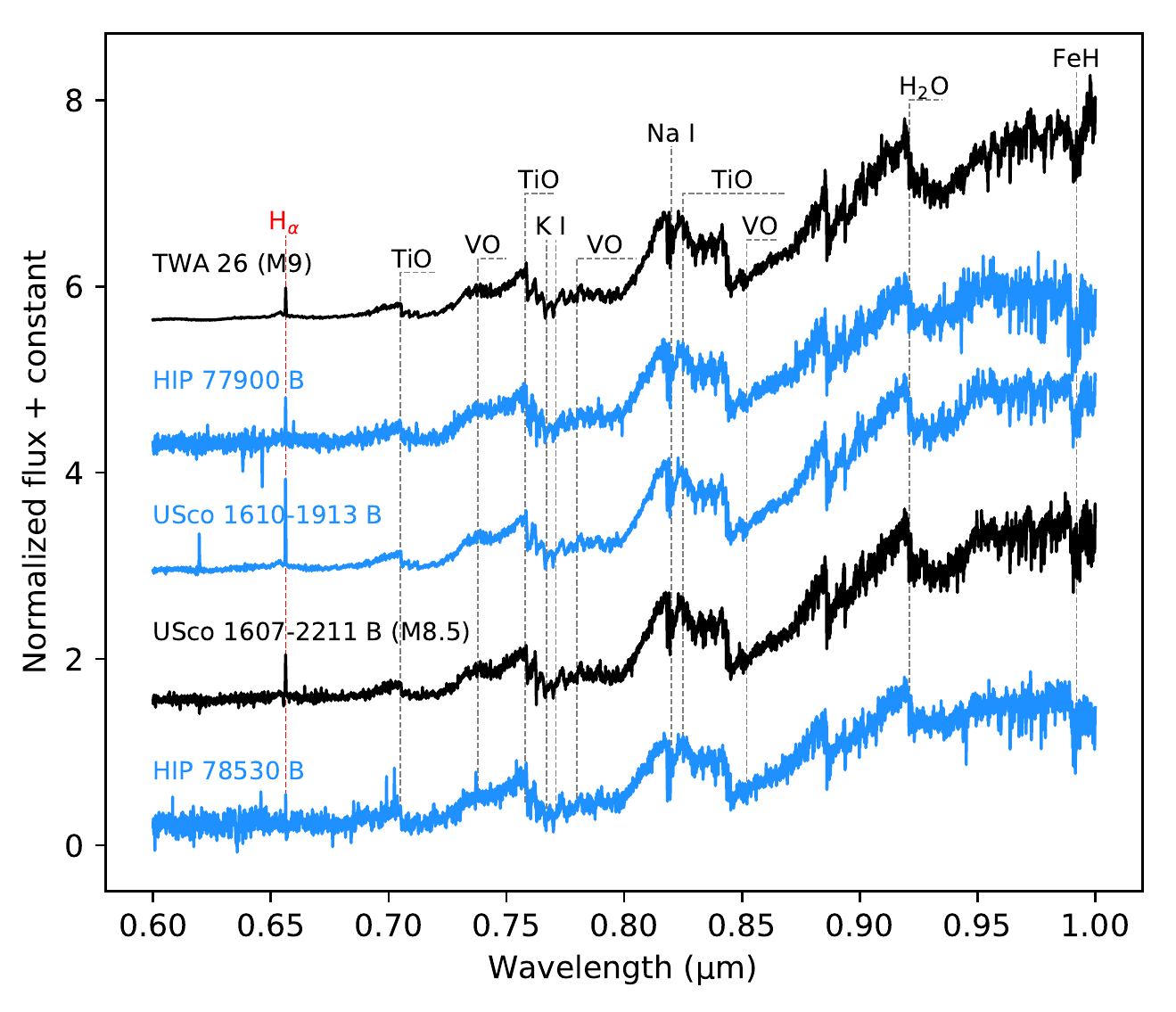} &  \includegraphics[width=8cm, angle=0]{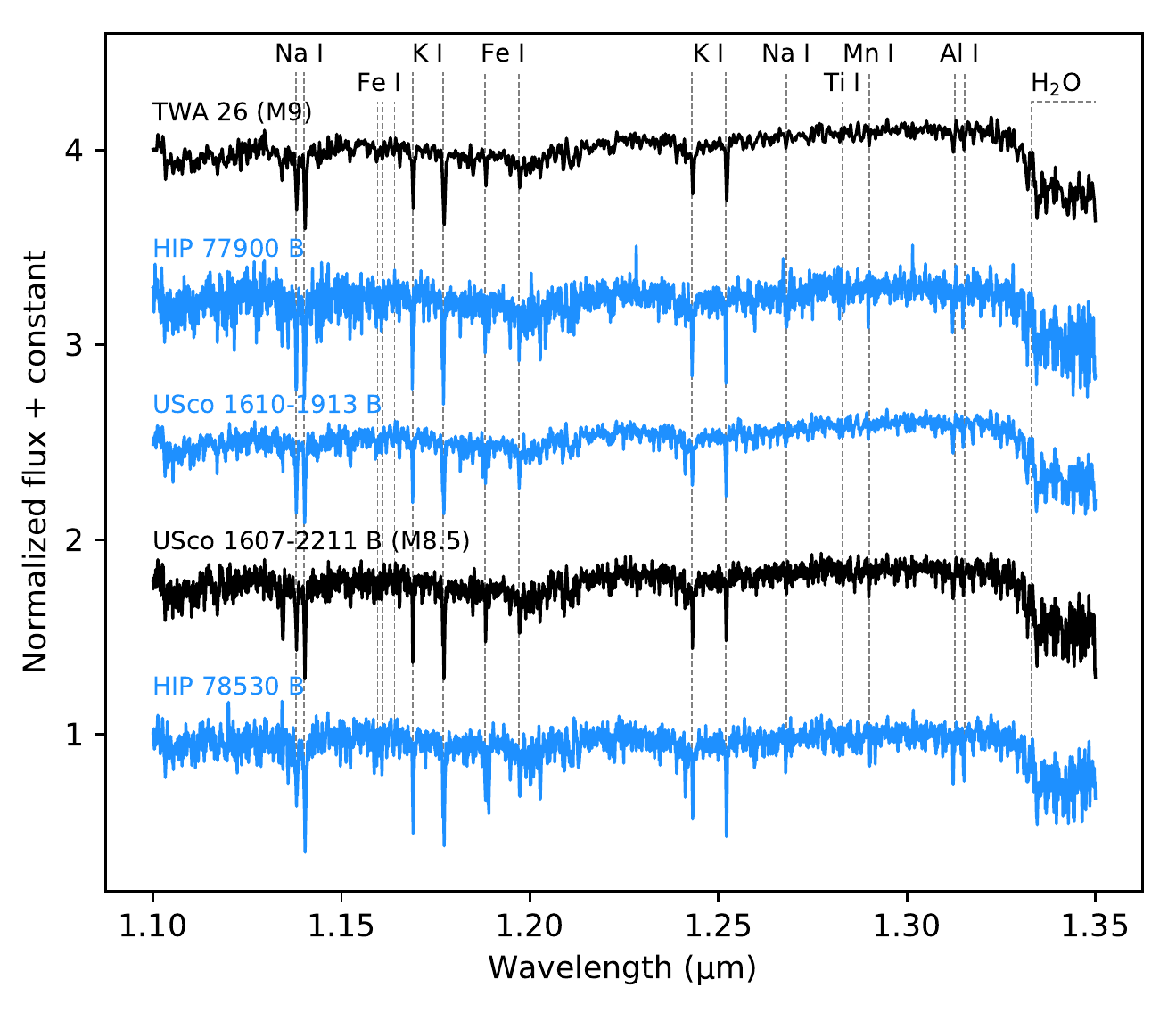} \\
   \end{tabular}
  \caption{Comparison of the optical spectra normalized at 0.82 \mic\, (\textit{left}) and J-band normalized at 1.32 \mic\, (\textit{right}) of USco\,1610-1913\,B, HIP\,77900\,B and HIP\,78530\,B (\textit{blue}) to the free-floating object USco\,1607-2211 from our original sample of spectra of Upper-Scorpius brown-dwarfs and to the TW Hydrae association member TWA\,26 \citep{2013AeA...551A.107M} (\textit{black}). All objects have a \Ha\,emission line (656.3nm; \textit{red label}) in addition to molecular and atomic absorption lines typical of late-M dwarfs.}
  \label{Fig:LineOptJ}
  \end{figure*}
 
\begin{figure*}[t]
  \centering
  \begin{tabular}{cc}
  \includegraphics[width=8cm, angle=0]{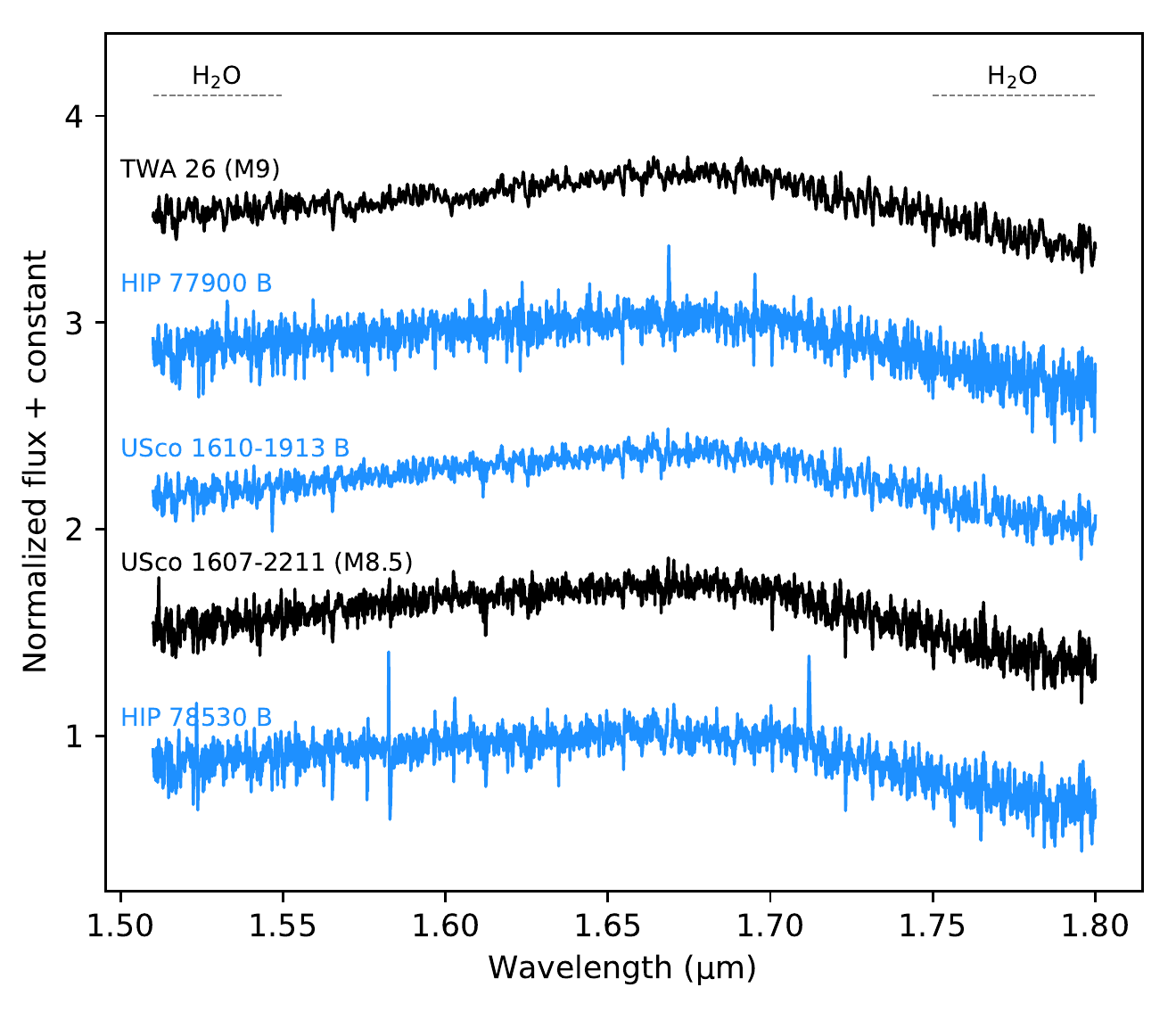} &  \includegraphics[width=8cm, angle=0]{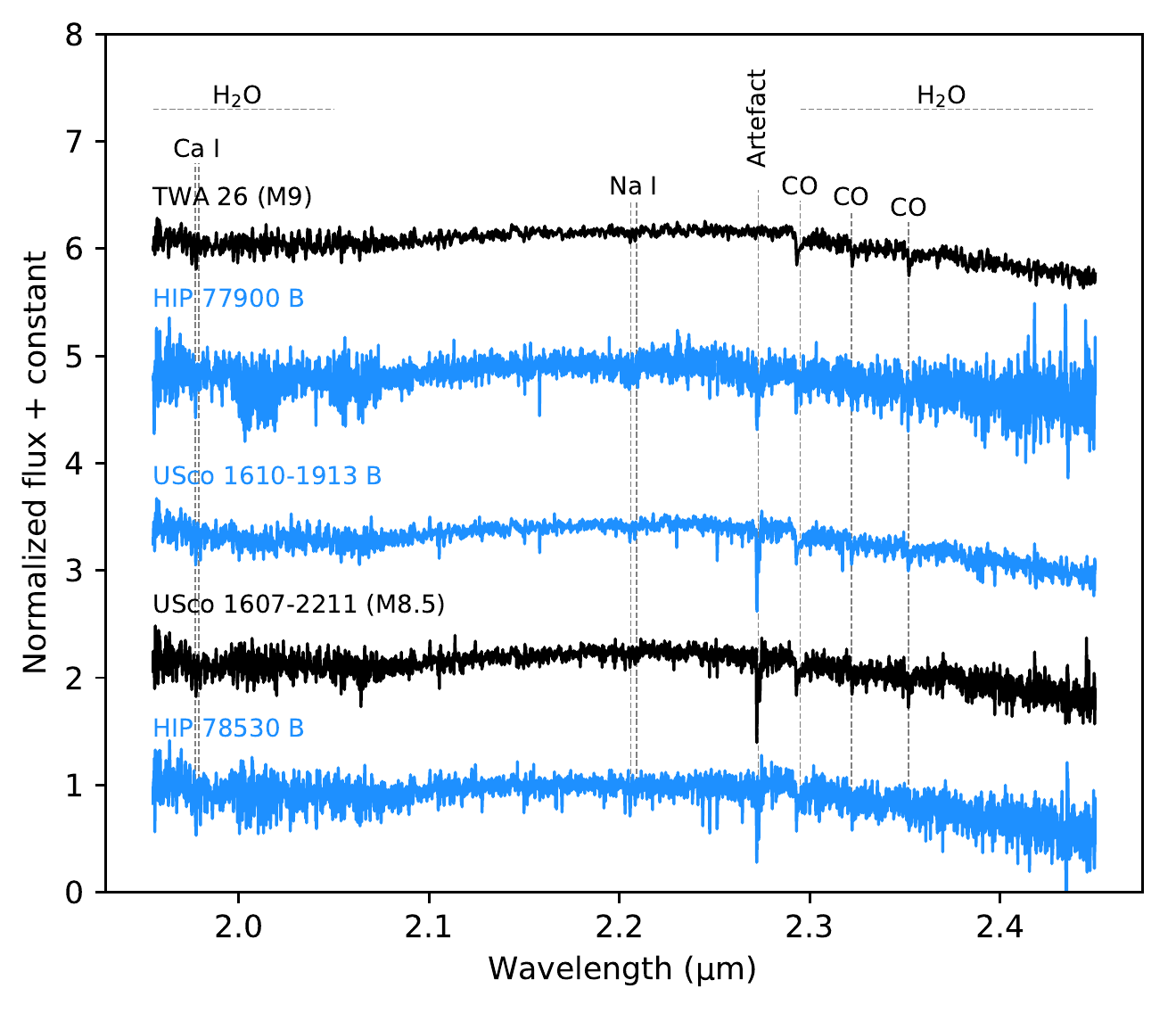}\\
\end{tabular}
  \caption{The same as Figure \ref{Fig:LineOptJ} but for H-band normalized at 1.65 \mic\, (\textit{left}) and K-band normalized at 2.15 \mic\, (\textit{right}).}
  \label{Fig:LineID_HK}
\end{figure*}

Figures \ref{Fig:LineOptJ} and \ref{Fig:LineID_HK} show the UVB+optical, J-band, H-band and K-band segments of the X-Shooter spectra of HIP\,78530\,B, HIP\,77900\,B, and USco\,1610-1913\,B together with two young, isolated brown dwarfs USco\,1607-2211 (M8.5) and TWA\,26 (M9). In the optical part, one can easily identify in Figure \ref{Fig:LineOptJ} (\textit{left}) the detection of the strong \Ha\, line (0.6563 \mic). Additional Balmer lines, \Hg\, (0.4340 \mic) and \Hb\, (0.4861 \mic), are also detectable in the X-Shooter spectra of USco\,1610-1913\,B and HIP\,77900\,B (also HIP\,78530\,B) indicating the possible presence of accretion or chromospheric activity. The Doublet of Ca II-H/K (0.3934, 0.3969 \mic) is also detected. The detection of these emission lines is discussed in Section \ref{sec:Emission line properties}. 

In J-band we identified several strong absorption lines, like the neutral sodium (Na I) doublet (1.138 and 1.141 \mic), the neutral potassium (K I) doublets (1.168, 1.177 \mic\, and 1.243, 1.254 \mic). Neutral iron (Fe I) lines (1.189 and 1.197 \mic) are also present as well as weaker Na I (1.268 \mic), Magnesium (Mn I at 1.290 \mic) and titanium (Ti I at 1.283 \mic) lines. We notice the typical triangular shape of the H-band which is produced by wide H$_{2}$O absorption bands and testifies to the young ages of our objects. In K-band we found the calcium (Ca I) triplet (at 1.98 \mic) and a weak Na I doublet (2.206, 2.209 \mic). We also detect CO bands (2.295, 2.322, 2.352 \mic).

\subsubsection{Spectral type and surface gravity  determination}
\begin{figure}
  \centering
  \includegraphics[width=\columnwidth, angle=0]{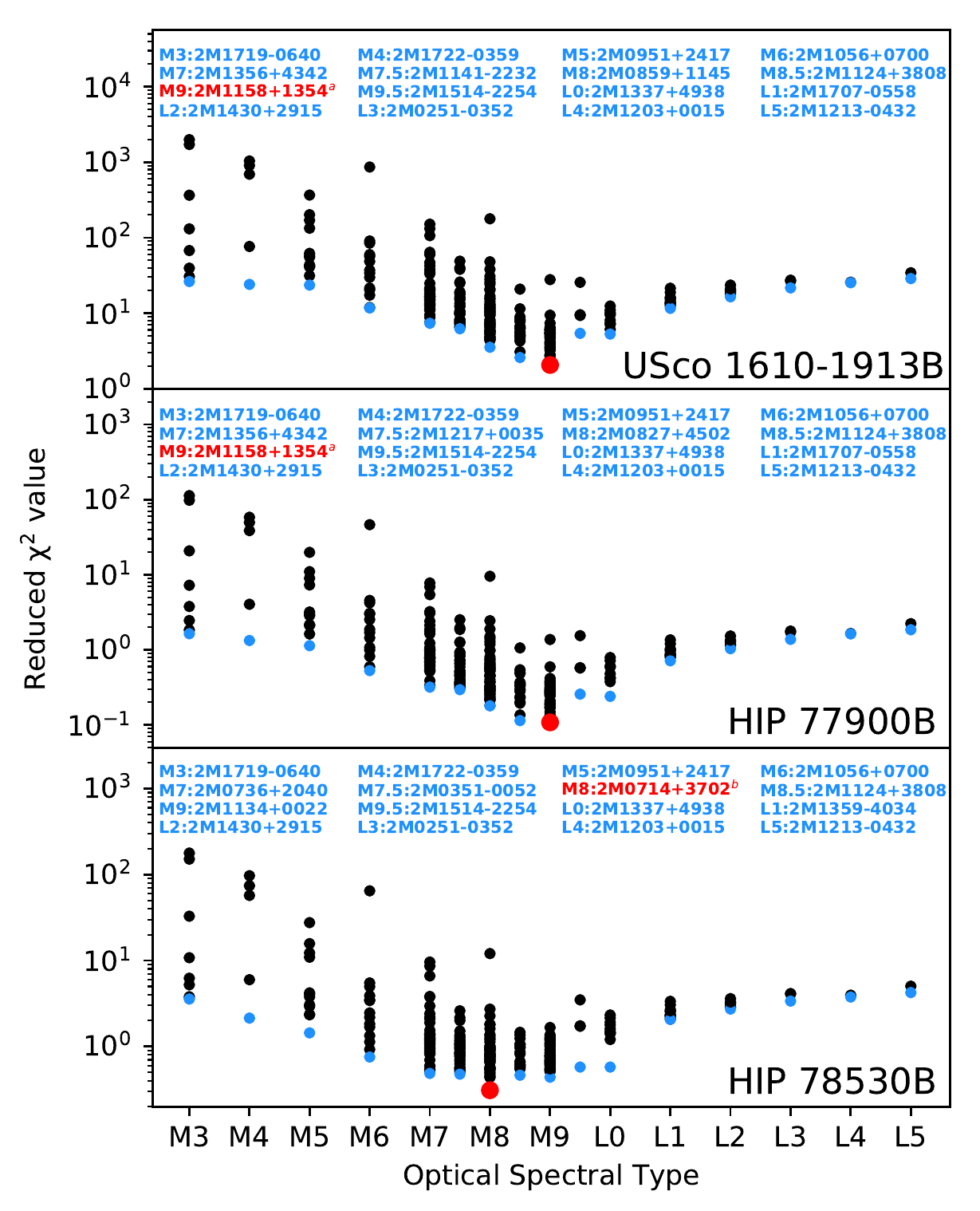}
  \caption{$\mathrm{\chi}^{2}$ from the comparison between the X-Shooter spectra of USco\,1610-1913\,B, HIP\,77900\,B, and HIP\,78530\,B to the optical spectra of Ultracool RIZzo spectral library. We indicate the best fitting object for each group in \textit{blue}. The $\mathrm{\chi}^{2}$ minimum for all objects is represented in \textit{red}. (a) : \cite{2003AJ....126.2421C}. (b) : \cite{2008AJ....136.1290R}}
  \label{Fig:khi2}
\end{figure}

We re-investigated the optical classification of our targets using a standard $\mathrm{\chi^{2}}$\, comparison of our spectra to empirical templates from the Ultracool RIZzo Spectral Library\footnote{https://jgagneastro.com/the-ultracool-rizzo-spectral-library/}. The RIZzo library is made of spectra of 265 M2-L5 brown dwarfs from \cite{2003AJ....126.2421C}, \cite{2007AJ....133..439C} and \cite{2008AJ....136.1290R}. 
We restrained the fit to the 0.75$\mic$\,\-\,0.86$\mic$\, range. 
The results of our spectral fitting are shown in Figure \ref{Fig:khi2} and are reported in Table \ref{tab:emp_resu}. The spectral type errors are due to the sub-group increment defined in the library. For USco\,1610-1913\,B, HIP\,77900\,B and HIP 78530\,B, we find spectral types of M$9\,\pm\,0.5$, M$9\,\pm\,0.5$, and M8$\,\pm\,0.5$, respectively with best fits 2MASS\,J11582484+1354456 (M9) and 2MASS\,J07140394+3702459 (M8) which are both free-floating objects (determined from \texttt{BANYAN $\Sigma$ Tool}). We also classified our targets in the near-infrared using absorption lines respecting the \cite{2013ApJ...772...79A} scheme. In near-infrared, we find later spectral types for USco\,1610-1913\,B and HIP 78530\,B than the ones derived by \cite{2015ApJ...802...61L}  consistent with the redder slope of the X-Shooter spectrum of HIP 78530\,B, and because of the revised extinction $\mathrm{A_{v}}$\, values (see also Section \ref{Sec:Results}) considered for the two systems. The optical spectral types derived for the free-floating BDs are in agreement with the ones of \cite{2018MNRAS.473.2020L} within error bars. 

Appendix \ref{sec:AppB} shows the systematic differences between each method. The spectral type derived from the H$_{2}$O index seems to be 1-2 subtype over the one from the visual method. \cite{2013ApJ...772...79A} explain that the H$_{2}$O index could be sensitive to gravity and so can be biased. We choose the spectral type from the visual comparison to avoid this bias.
 
In addition to the spectral-type determination, we also applied the surface gravity classification formalized by \cite{2013ApJ...772...79A}. The results are reported in Table \ref{tab:emp_resu} and show that HIP\,77900\,B and HIP\,78530\,B are identified as young, intermediate surface gravity BDs. USco\,1610-1913\,B is confirmed as a very-low surface gravity BD. The gravity class of  HIP\,78530\,B is consistent with the one derived by \cite{2015ApJ...802...61L}. Our results are consistent with those of \cite{2018MNRAS.473.2020L} for the classification of the young free-floating objects.

\subsubsection{Over-luminosity of USco\,1610-1913\,B}
Figure \ref{Fig:Overlum} shows the comparison between the visible and near-infrared calibrated flux of USco\,1610-1913\,B with the fluxes of young reference BDs of similar spectral types from our sample with \textit{Gaia} parallaxes.  We also include a comparison to the X-Shooter spectrum of TWA\,26 \citep{2013AeA...551A.107M}. All BDs have been scaled to the distance of USco\,1610-1913A. All these objects reproduce the detailed absorptions  and pseudo-continuum shape of USco\,1610-1913\,B spectrum provided that an extra scaling factor of 2 (TWA\,26) and 4 (USco\,1610-2239, HIP\,77900\,B) is considered. This over-luminosity of USco\,1610-1913\,B  had already been noted by \cite{2013ApJ...773...63A} and \cite{2015ApJ...802...61L}. We confirm it over our extended wavelength range and resolution relying on an extended set of comparison objects from the association with now published parallaxes. We  discuss the possible origins of the over-luminosity in Section \ref{Sec:Discussion}.

 \begin{table*}[t]
\caption{Results of the empirical analysis}
\label{tab:emp_resu}
\renewcommand{\arraystretch}{1.3}
\small
\begin{center}
\begin{center}
  \begin{tabular}{c||c|ccc||cccc|c}
\hline
\hline
  &  \multicolumn{4}{c||}{Spectral Types}  &  \multicolumn{5}{c}{Gravity} \\
\hline
 Source  &  Visual  &     \multicolumn{3}{c||}{NIR}    &  \multicolumn{4}{c|}{Score\tnote{c}}  &  Type  \\
 \cline{3-9}
  &    &  H$_{2}$O (H)\tnote{a}  &  H$_{2}$O-1 (J)\tnote{b}  &  H$_{2}$O-2 (K)\tnote{b}  &  FeH$_{z}$  &  FeH$_{j}$  &  K I$_{j}$  &  H-cont  &  \\
\hline	
USco\,1610-1913\,B  & M9$\pm$ 0.5 & M9.1$\pm$ 0.2 & L0.1$\pm$ 0.2 & M8.9$\pm$ 0.2 & 2 & 2 & 2 & 2 & v-low \\
HIP\,77900\,B  & M9$\pm$ 0.5 & M8.4$\pm$ 0.5 & M9.9$\pm$ 0.4 & M9.5$\pm$ 0.5 & 1 & 2 & 1 & 2 & int \\
HIP\,78530\,B  & M8$\pm$ 0.5 & M8.4$\pm$ 0.4 & M8.4$\pm$ 0.4 & M8.2$\pm$ 0.4 & 1 & 2 & 1 & 2 & int \\
\hline
USco\,1607-2242  & M9$\pm$ 0.5 & L1.1$\pm$ 1.9 & L1.7$\pm$ 1.5 & L0.7$\pm$ 1.7 & n & 1 & 2 & 2 & v-low \\
USco\,1608-2232  & M9$\pm$ 0.5   & L0.9$\pm$ 0.8 & L1.5$\pm$ 0.6 & L0.1$\pm$ 0.7 & 2 & 2 & 1 & 2 & v-low \\
USco\,1606-2335  & M9$\pm$ 0.5   & M9.9$\pm$ 0.9 & L0.5$\pm$ 0.8 & M9.3$\pm$ 1.0 & 2 & 1 & 1 & 2 & int \\
USco\,1610-2239  & M9$\pm$ 0.5   & M9.9$\pm$ 0.5 & L0.3$\pm$ 0.4 & M9.3$\pm$ 0.5 & 2 & 2 & 2 & 2 & v-low \\
USco\,1608-2315  & M8.5$\pm$ 1.0 & L0.0$\pm$ 0.4 & L0.6$\pm$ 0.4 & M9.1$\pm$ 0.4 & 2 & 2 & 1 & 2 & v-low \\
USco\,1607-2211  & M8.5$\pm$ 1.0   & M9.4$\pm$ 0.5 & L0.4$\pm$ 0.4 & M8.3$\pm$ 0.5 & 2 & 2 & 1 & 2 & v-low \\
\hline
\hline
\end{tabular}
\end{center}
\begin{tablenotes}
\item Notes: relations and coefficients : [a] From \cite{2007ApJ...657..511A}, [b] From \cite{2004ApJ...610.1045S}, and [c] See \cite{2013ApJ...772...79A}. The systematic differences about the spectral type determination are discussed in Appendix \ref{sec:AppB}.
\end{tablenotes}
\end{center}
\end{table*}

	\subsection{Forward modelling analysis}
   	    \subsubsection{Description of the atmospheric models}
 We used a forward-modeling approach to determine the atmospheric parameters of the young BDs observed in this study. Forward modelling codes enable comparison of the object spectrum to pre-computed grids of models  which include our best knowledge of atmospheric physics.
       \label{BT-SETTL models}
We used grids of synthetic spectra produced by the \texttt{BT-SETTL15}\, model \citep{2012RSPTA.370.2765A}. This model handles the radiative transfer using the \texttt{PHOENIX} code \citep{1997ApJ...483..390H, 2001ApJ...556..357A}. It accounts for convection using the mixing-length theory, and works at hydrostatic and chemical equilibrium. The opacities are treated line by line (details on each elements are given in \citealt{2018AeA...620A.180R}). The code models the condensation, coalescence, and mixing of 55 types of grains. The abundances of solids are determined comparing the timescales of these different processes at each layer. In this study, we considered the predictions of the \texttt{BT-SETTL15}\, model with \Teff\, ranging from 2100\,K to 3000\,K (in steps of 100\,K), and a range of log(g) from 3.5 to 5.5\,dex (in steps of 0.5\,dex). We assumed a solar metallicity M/H\,=\,0.0, in broad agreement with the values reported in Sco-Cen \citep{2011AJ....142..180B}. These grids have been extensively utilized in previous studies of young BDs \citep{2014AeA...564A..55M, 2014AeA...562A.127B, 2017AeA...608A..79D, 2017MNRAS.465..760B}, but never as part of a Bayesian methodology, as developed here with the \texttt{ForMoSA} code.
        
  \begin{figure}[t]
  \centering
  \includegraphics[width=\columnwidth, angle=0]{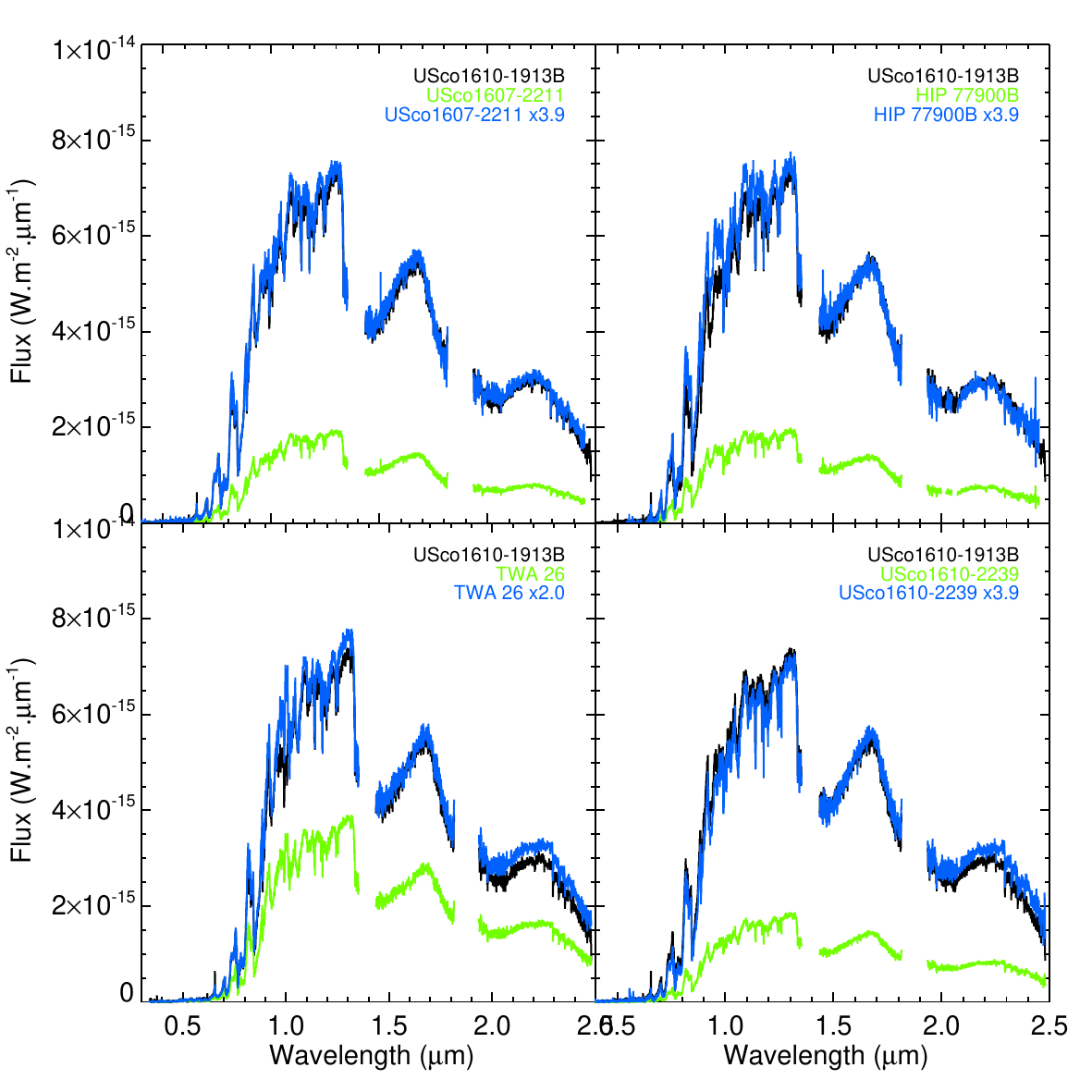}
  \caption{Comparison of the flux-calibrated 0.4-2.5 \mic\, spectra of USco\,1610-1913\,B (\textit{black}) to those of reference objects scaled to the  companion's distance (green). The companion is 2.0 to 3.9 times more luminous than objects having the same spectral type.}
  \label{Fig:Overlum}
\end{figure}

	    \subsubsection{The \texttt{ForMoSA} code}
	   For this work, we chose to develop our own forward modelling code relying on the Nested Sampling procedure \citep{skilling2006}. The  method explores and recursively isolates different patches of likelihood levels in the parameter space. It offers several advantages over classical Markov-Chain-Monte-Carlo algorithms also used for the forward modelling (e.g. \citealt{2018AeA...618A..63B,2017AeA...603A..57S}). The approach avoids missing local minimum within vast and degenerate parameter spaces while ensuring the convergence of the exploration.
        
	    The code \texttt{ForMoSA} (for FORward MOdeling for Spectral Analysis) takes as input an observed spectrum with associated error bars and any grid of synthetic spectra (here the \texttt{BT-SETTL15} grid). The Nested Sampling is handled by the \textit{nestle} Python module\footnote{http://kylebarbary.com/nestle/}. \texttt{ForMoSA} re-samples the data and the models to make them comparable. The grid of model spectra are first interpolated onto the wavelength grid of the observation and degraded to the spectral resolution of the instrument that acquired the data. The optical spectra of our targets and the models were degraded to R$_{\lambda}\,=\,3300, 5400$ and $4300$, similar to the UVB, VIS and NIR part of the spectrum respectively. 
        
        In our case, we suppose independent data points in the observed spectra, so the likelihood is derived from the $\mathrm{\chi^{2}}$\, value. To compute $\mathrm{\chi^{2}}$\, at each step, we need to generate a model spectrum for a set of free-parameters that does not necessarily exist in the original grid of spectra. Therefore, we generated a model spectrum on-demand following a two-step process: \begin{itemize}
        \item A preliminary phase consists in reducing the grid meshes. To do so, \texttt{ForMoSA} interpolates and reduces each \Teff\, increment to 10\,K and log(g) increments to 0.01\,dex.  We have considered the linear and bicubic spline interpolation approaches, and finally selected the bicubic spline interpolation which better accounts for the flux variation through the grid. This step needs to be done one time and ensure a regular grid. In doing this, we increase the accuracy of the second interpolation phase.
        \item The second phase arises in the course of the Nested Sampling process when a new point in the parameter space is defined. The closest neighbours approach was found to provide the best trade-off between the reliability of the interpolation process and the computation time needed to run the interpolation.
        \end{itemize}
        
\begin{figure}
  \centering
  \includegraphics[width=\columnwidth, angle=0]{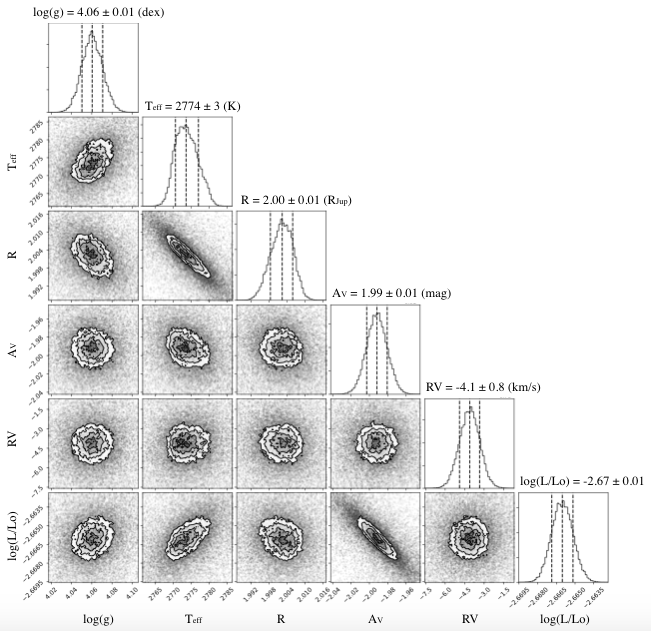}
  \caption{Posteriors of HIP\,78530\,B in using \texttt{ForMoSA} on the $J+H+K$ and with the extinction parameter free.}
  \label{Fig:corner_HIP_78530B_Av}
\end{figure}

        Each synthetic spectrum gives the flux at the top of the atmosphere. The comparison with the observed spectrum requires in addition multiplying the model by a dilution factor $C_{k}\,=\,(\frac{R}{d})^{2}$, with R the object radius and d the distance. We adopt distances for the host stars from the \textit{Gaia} DR2. We considered flat priors on \Teff\, (2100-2900\,K) and log(g) (3.5-5.5\,dex). \Teff\, and radius are linked together by the luminosity (Boltzmann law) so we have chosen a flat prior on R (0.5-30.0\,\RJup) to be conservative and ensure we do not limit the exploration of \Teff. We also provide the luminosity from the posterior distributions of $R$ and \Teff\, \texttt{ForMoSA} can also consider interstellar extinction ($\mathrm{A_{v}}$) as a free parameter using the extinction law from \cite{2007ApJ...663..320F}, the radial velocity (RV) by a Doppler shifting law\footnote{function PyAstronomy.pyasl.dopplerShift} and the projected rotational velocity v.sin(i) according to the rotational broadening law from Gray’s "The Observation and Analysis of Stellar Photospheres"\footnote{function PyAstronomy.pyasl.rotBroad}. Figure \ref{Fig:corner_HIP_78530B_Av} illustrates a typical output of \texttt{ForMoSA} on HIP\,78530\,B. The method allows us to identify correlations between parameters such as \Teff\,and R.
        
	\subsection{Results}
	\label{Sec:Results}

\begin{figure}
  \centering
  \includegraphics[width=\columnwidth, angle=0]{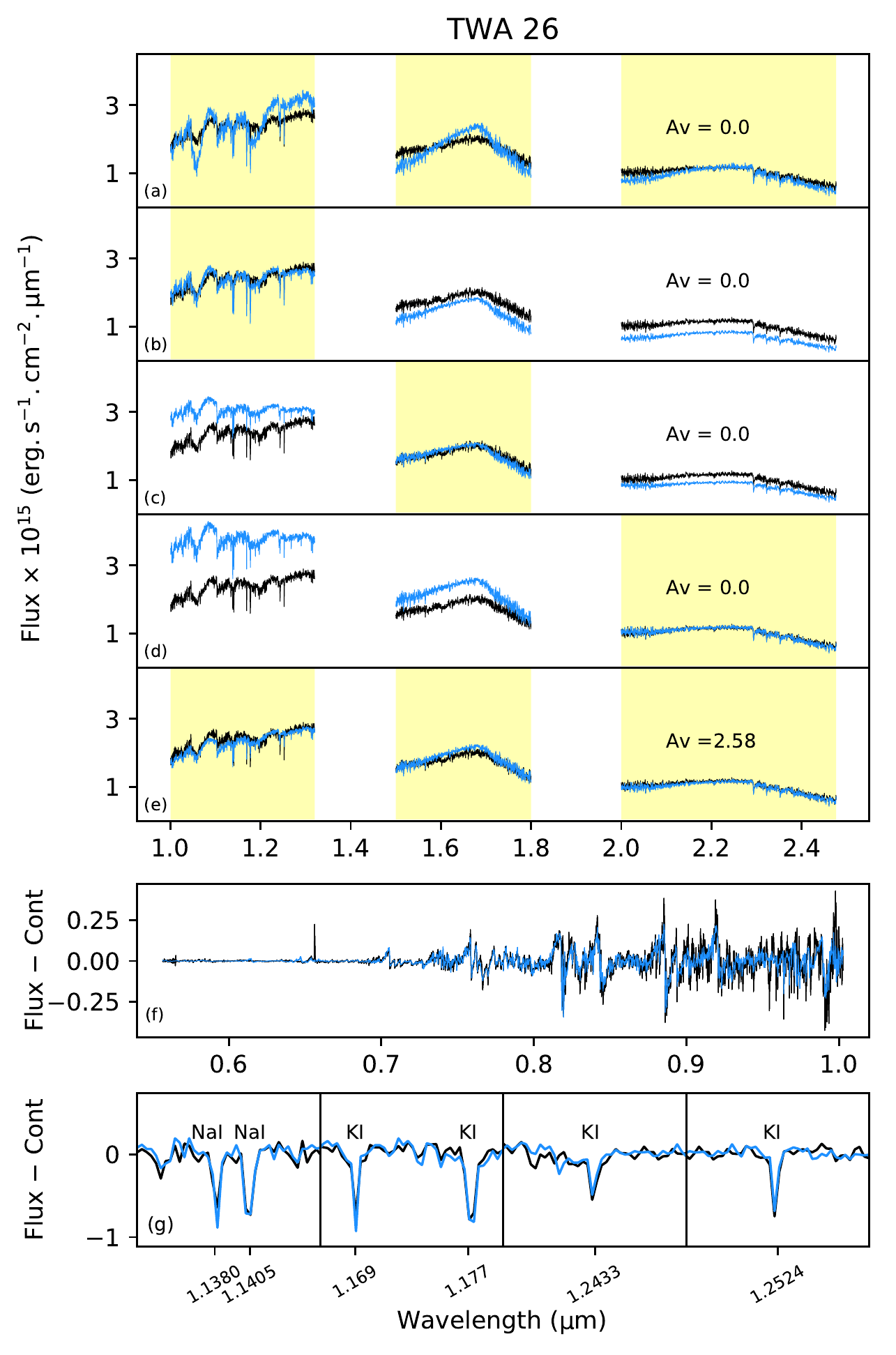}
  \caption{Best fits of TWA\,26 (in \textit{black}) by \texttt{BT-SETTL15}\, models (in \textit{blue}) using different portions of the spectrum for the fit (shaded \textit{yellow} rectangles). We define the fitting zones as follow : $J\,=\,1.0-1.32$ \mic\, ; $H\,=\,1.5-1.8$ \mic\, ; $K\,=\,2.0-2.48$ \mic. We perform the following fits : (a) $J+H+K$ ; (b) $J$ ; (c) $H$ ; (d) $K$ ; (e) $J+H+K$ with Av as a free parameter ; (f) the optical part ($0.56-1.00$ \mic) on the flux-continuum and v.sin(i) as a free parameter ; (g) examples of lines fitted on the flux-continuum.}
  \label{Fig:diff_fit_TWA26}
\end{figure}

\begin{figure}
  \centering
  \includegraphics[width=\columnwidth, angle=0]{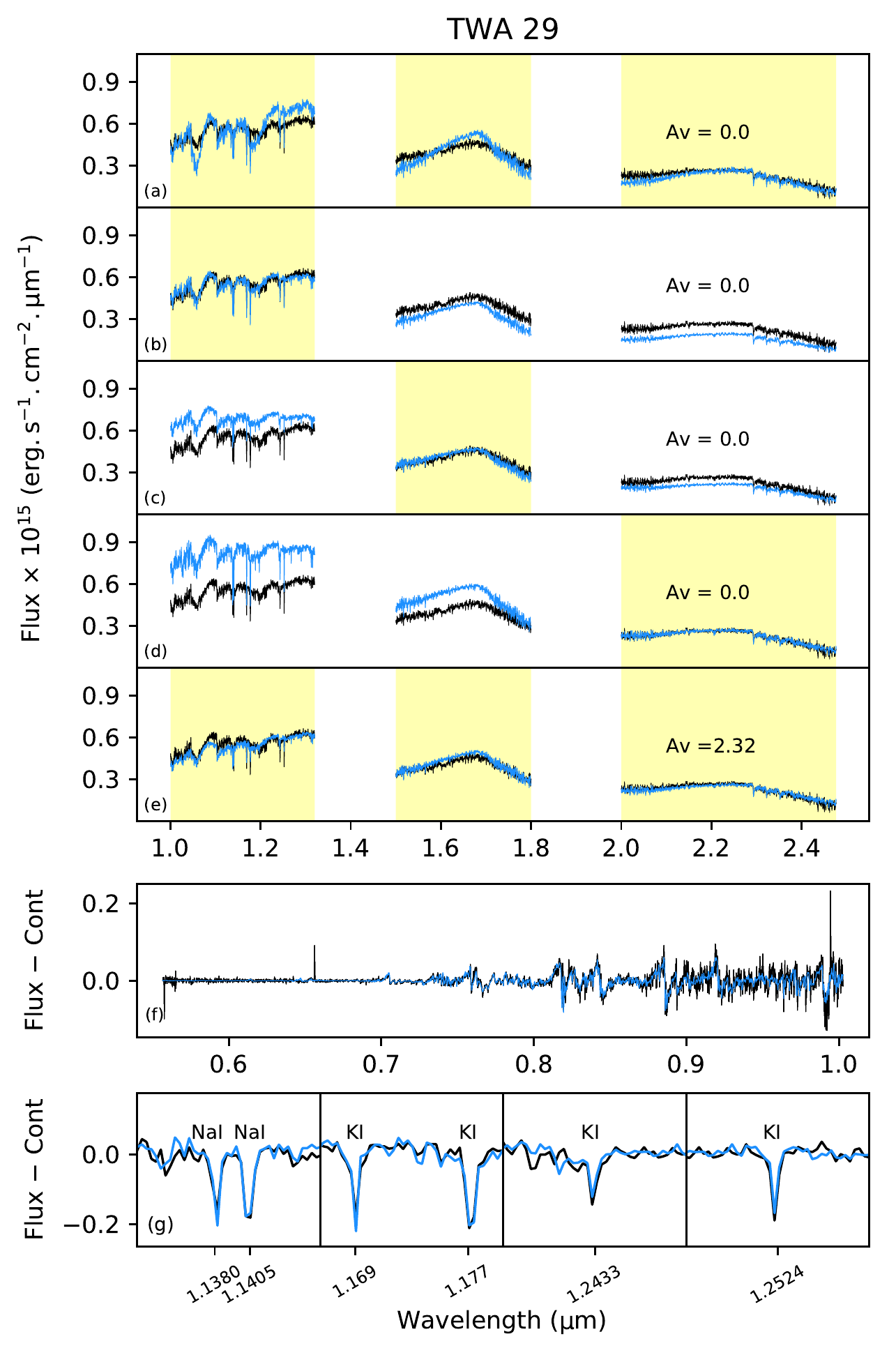}
  \caption{Same as Figure \ref{Fig:diff_fit_TWA26} but for TWA\,29.}
  \label{Fig:diff_fit_TWA29}
\end{figure}

\begin{figure}
  \centering
  \includegraphics[width=\columnwidth, angle=0]{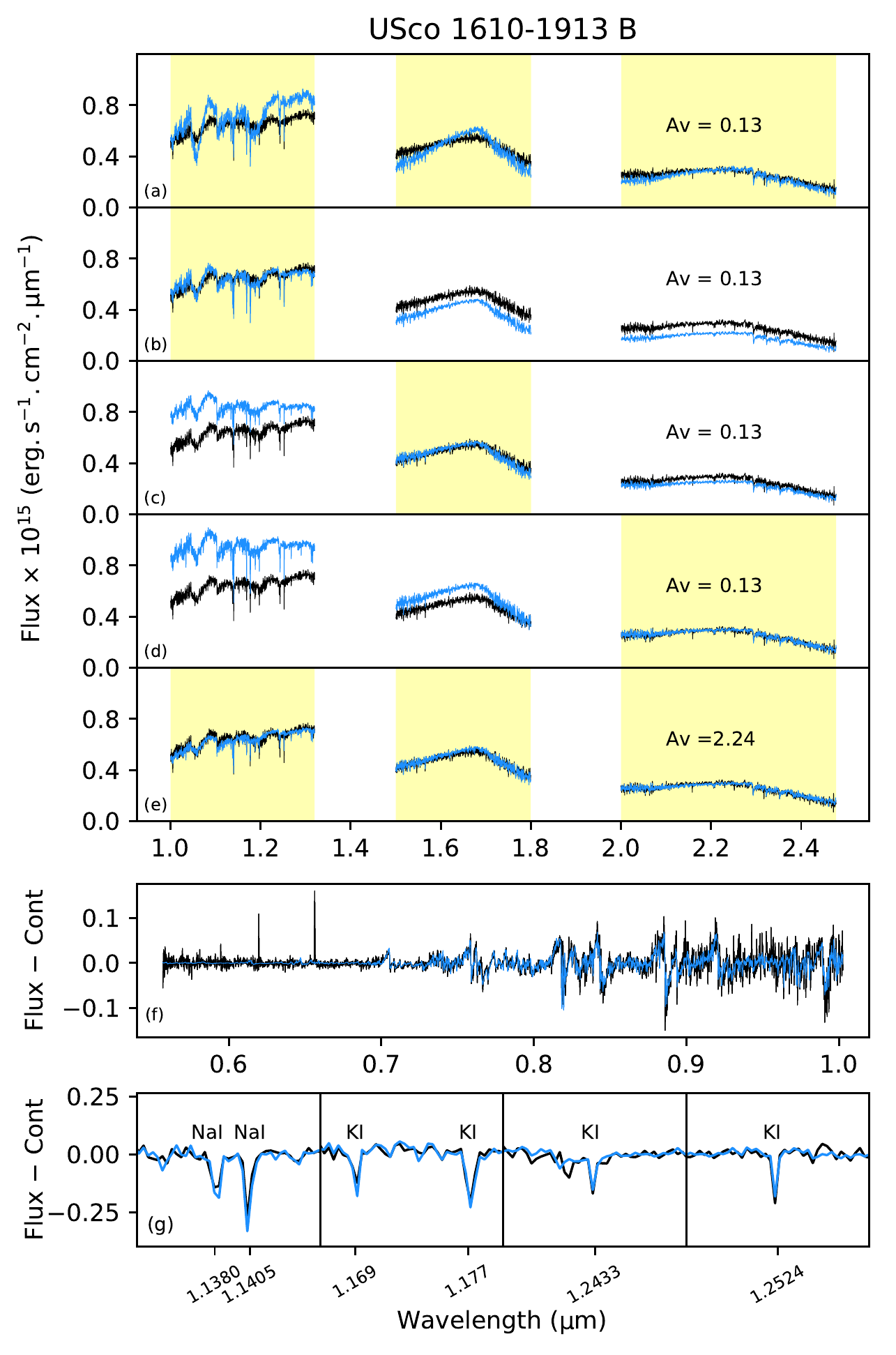}
  \caption{Same as Figure \ref{Fig:diff_fit_TWA26} but for USco\,1610-1913\,B.}
  \label{Fig:diff_fit_USco1610_1913B}
\end{figure}

\begin{figure}
  \centering
  \includegraphics[width=\columnwidth, angle=0]{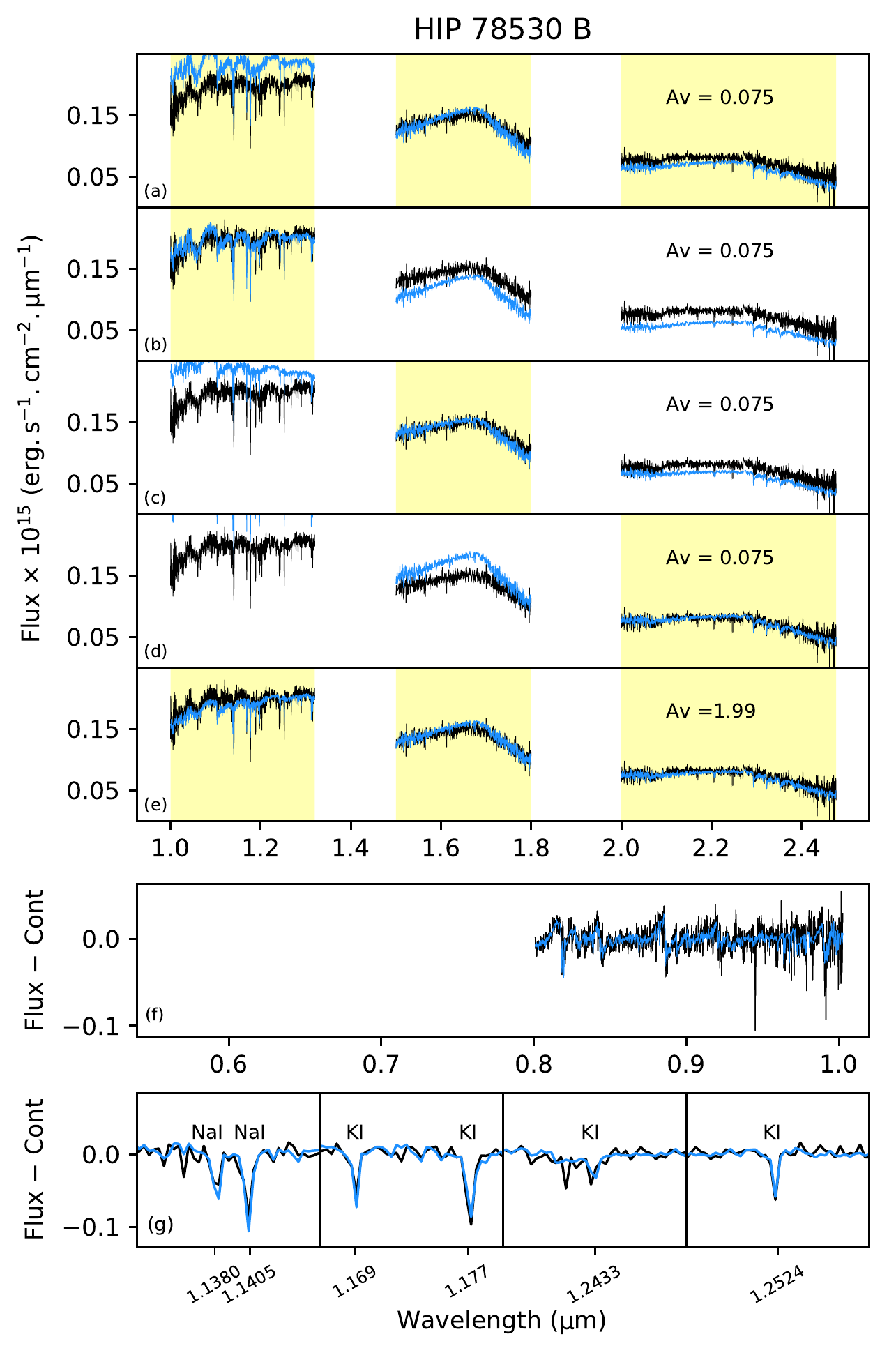}
  \caption{Same as Figure \ref{Fig:diff_fit_TWA26} but for HIP\,78530\,B. We avoid the flux inconsistency presented in Section \ref{Sec:Empirical_analysis} in using the wavelength range 0.8-1.0\,\mic\,in panel (f).}
  \label{Fig:diff_fit_HIP78530B}
\end{figure}

\begin{figure}
  \centering
  \includegraphics[width=\columnwidth, angle=0]{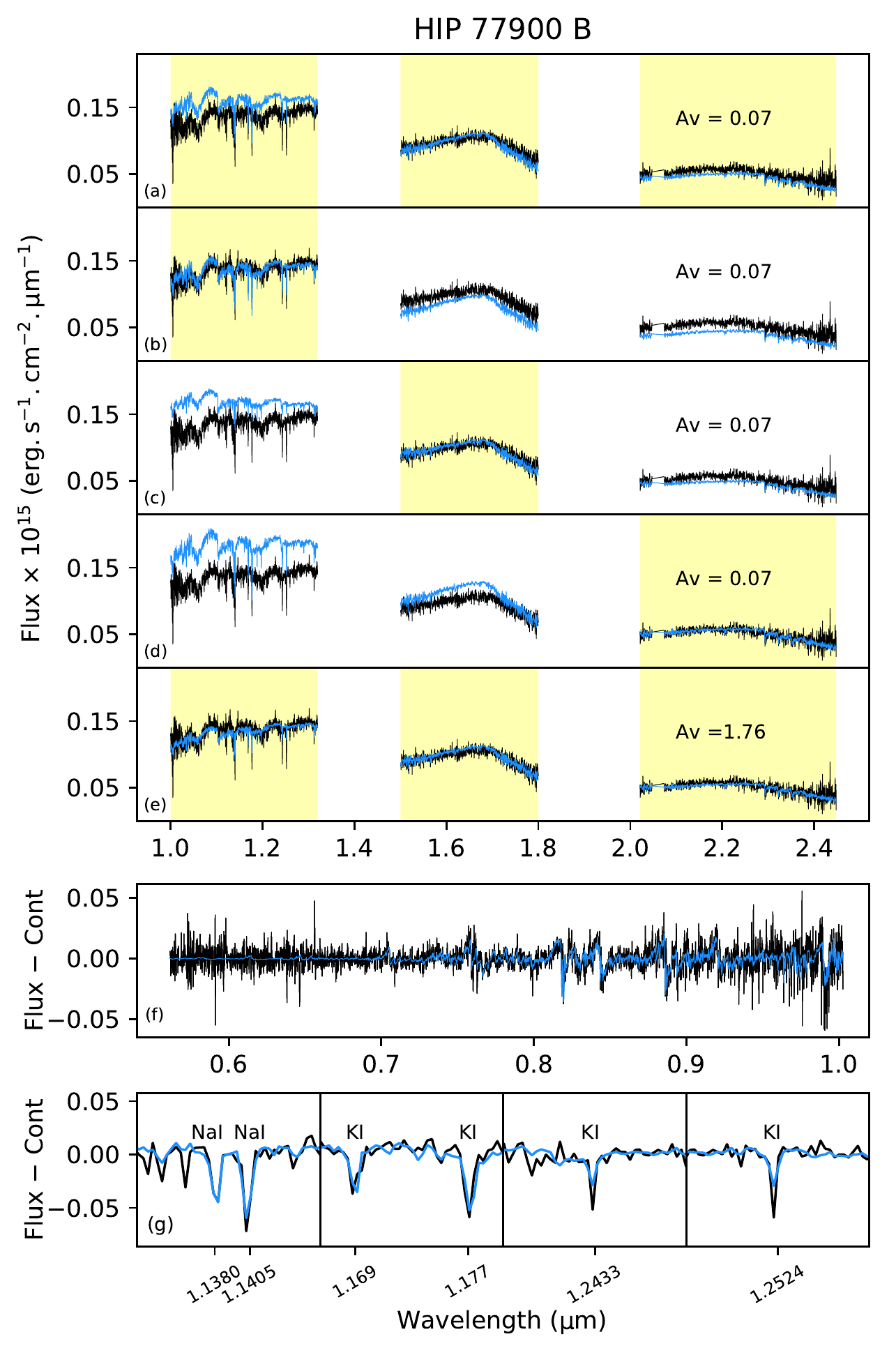}
  \caption{Same as Figure \ref{Fig:diff_fit_TWA26} but for HIP\,77900\,B.}
  \label{Fig:diff_fit_HIP77900B}
\end{figure}

   \begin{table}[t]
\caption{Physical properties of TWA\,26 and TWA\,29.}
\label{Tab:comparison_TWA_Fil}
\renewcommand{\arraystretch}{1.8}
\small
\begin{tabular}{c|cccc|c}
\hline
\hline
& \Lum	&	R  &  log(g) & \Teff  	\\
&    	&	(\RJup)  &  (dex) & (K)  & \tiny Ref 	\\
\hline
\multirow{2}*{\rot{{\tiny TWA\,26}}}
& $-2.71\pm0.09$ & $2.20\pm0.22$ & $4.14\pm0.16$ &  $2552\pm188$    & \tiny \textit{a} \\
& $-2.83^{+0.38}_{-0.36}$ & $1.97^{+0.87}_{-0.52}$ & $\leq\,4.11$ & $2547^{+94}_{-136}$ &  \tiny \textit{b} \\        
\hline
\multirow{2}*{\rot{{\tiny TWA\,29}}}
& $-2.79\pm0.15$ & $2.00\pm0.23$ & $4.13\pm0.15$ &  $2391\pm249$    & \tiny \textit{a} \\
& $-2.77^{+0.10}_{-0.11}$ & $2.14^{+0.15}_{-0.10}$ & $\leq\,4.33$ & $2522^{+58}_{-99}$ &  \tiny \textit{b} \\          \hline
\hline
\end{tabular}
\tablefoot{\textit{a} \cite{2015ApJ...810..158F}, \textit{b} this work with the adopted values (See Table \ref{Tab:Results_FORMOSA1}-\ref{Tab:Results_FORMOSA2}).}
\end{table}

\begin{table*}
 \centering
 \begin{threeparttable}
\renewcommand{\arraystretch}{1.3}

  \caption{Comparison between parameters from \texttt{ForMoSA} and parameters in using evolutionary models \citep{2000ApJ...542L.119C} (\texttt{BT-SETTL15}/\texttt{ForMoSA} independent). The  luminosity has been determined using the bollometric corrections at J and K bands from \cite{2015ApJ...810..158F}. For the luminosity of our reference objects we directly use the \cite{2015ApJ...810..158F} values. We also provide the mass values.}
\tiny   
  \begin{tabular}{c|cccc||cccc|c}
  \hline   
	\hline
 & \multicolumn{4}{c||}{\texttt{ForMoSA}}  & \multicolumn{5}{c}{Bolometric correction + Evolutionary models} \\	
Object			         & \Lum                    & R                      & log(g)       & \Teff              &  \Lum      & R           & log(g)        & \Teff  &    M	  \\
				         &                         & (\RJup)                & (dex)        &  (K)               &     & (\RJup) &  (dex)        & (K) &(\MJup) 	\\
\hline
USco\,1610-1913\,B \footnotemark[1]      & $-2.25^{+0.10}_{-0.10}$ & $3.87^{+0.24}_{-0.12}$ & $\leq\,4.17$ & $2542^{+68}_{-104}$  & $-2.13\pm0.17$ & $3.6\pm0.7$ & $4.03\pm0.20$ & $2827\pm169$ & $57\pm28$ \\
HIP\,77900\,B \footnotemark[1]	         & $-2.89^{+0.15}_{-0.13}$ & $1.76^{+0.15}_{-0.12}$ & $\leq\,4.36$ & $2602^{+117}_{-97}$ & $-2.59\pm0.16$ & $2.8\pm0.4$ & $4.02\pm0.17$ & $2611\pm185$ & $34\pm14$ \\
HIP\,78530\,B \footnotemark[1]	         & $-2.87^{+0.15}_{-0.15}$ & $1.83^{+0.16}_{-0.14}$ & $\leq\,4.34$ & $2679^{+118}_{-119}$  & $-2.68\pm0.17$& $2.6\pm0.4$ & $4.00\pm0.13$ & $2544\pm179$ & $28\pm10$ \\
\hline
USco\,1607-2242          & $-3.41^{+0.18}_{-0.18}$ & $1.13^{+0.13}_{-0.08}$ & $\leq\,4.10$ & $2403^{+122}_{-152}$ & $-3.23\pm0.20$ & $1.9\pm0.2$ & $3.99\pm0.09$ & $2044\pm203$ & $14\pm4$ \\
USco\,1608-2232          & $-3.09^{+0.15}_{-0.14}$ & $1.63^{+0.19}_{-0.12}$ & $\leq\,4.09$ & $2409^{+81}_{-99}$   & $-2.96\pm0.20$ & $2.1\pm0.3$ & $4.01\pm0.06$ & $2262\pm182$ & $17\pm4$ \\
USco\,1606-2335          & $-3.11^{+0.16}_{-0.16}$ & $1.46^{+0.15}_{-0.10}$ & $\leq\,4.36$ & $2519^{+114}_{-141}$  & $-3.05\pm0.20$ & $2.1\pm0.2$ & $4.01\pm0.06$ & $2215\pm179$ & $16\pm4$ \\
USco\,1610-2239          & $-2.88^{+0.17}_{-0.14}$ & $1.93^{+0.24}_{-0.14}$ & $\leq\,4.01$ & $2499^{+102}_{-108}$  & $-2.75\pm0.20$ & $2.5\pm0.3$ & $3.97\pm0.08$ & $2467\pm173$ & $24\pm7$ \\
USco\,1608-2315          & $-2.86^{+0.11}_{-0.15}$ & $2.00^{+0.12}_{-0.09}$ & $\leq\,4.16$ & $2487^{+81}_{-147}$ & $-2.71\pm0.20$ & $2.5\pm0.4$ & $3.98\pm0.08$ & $2474\pm179$ & $24\pm8$ \\
USco\,1607-2211          & $-2.84^{+0.07}_{-0.11}$ & $1.92^{+0.05}_{-0.06}$ & $\leq\,4.05$ & $2557^{+65}_{-117}$  & $-2.76\pm0.24$& $2.5\pm0.4$ & $3.98\pm0.09$ & $2461\pm204$ & $24\pm8$  \\
\hline
\hline
 \end{tabular}
\label{Tab:param_compare}
\begin{tablenotes}
\item [1] With the hypothesis that companions are at the same distance than the primary star.
\end{tablenotes}
\end{threeparttable}

\end{table*}

As a safety check, the \texttt{ForMoSA} code was applied to our three companions HIP\,78530\,B, HIP\,77900\,B, and USco\,1610-1913\,B and two well known young BDs, TWA 26 and TWA 29, with similar spectral types and ages ($\sim8\,$Myr) and observed by \cite{2013AeA...551A.107M}. We considered wavelengths from 1.0\,\mic\, to 2.5 \mic\, in all our fits to avoid biases related to the residual contamination of HIP\,78530\,B and run a homogeneous analysis on all objects.  The results are shown in Figures \ref{Fig:diff_fit_TWA26} - \ref{Fig:diff_fit_HIP77900B}. The numerical values are given in  Tables\,\ref{Tab:Results_FORMOSA1} - \ref{Tab:Results_FORMOSA2}.
	    
\begin{itemize}
\item Panels (a), (b), (c) and (d) of Figures \ref{Fig:diff_fit_TWA26} - \ref{Fig:diff_fit_HIP77900B} show the X-Shooter spectra (\textit{black}) with the resulting best fit (\textit{blue}) considering different spectral windows ($J+H+K$, $J$, $H$ and $K$, respectively), and a fixed-extinction assumption from the \textit{Gaia} DR2 extinction maps and values at the location of our targets \citep{2019arXiv190204116L}. From these maps, we calculate the $\mathrm{A_{0}}$ extinction at 550.0\,nm and assume that the difference with the $\mathrm{A_{v}}$ extinction is negligible ($\mathrm{A_{0}}$\,$\simeq$\,$\mathrm{A_{v}}$). The fitting spectral range is indicated by the \textit{yellow} background. 
\item Panel (e) shows the best fit when using the complete $J+H+K$-band spectral range, and adding the extinction value $\mathrm{A_{v}}$ this time as a free parameter. We reddened the models with an extinction function in the code \citep{2007ApJ...663..320F}. 
\item Panel (f) shows the best fit focused on the optical part (0.56-1.00 \mic\,for all targets but 0.80-1.00 \mic\,for HIP\,78530\,B). We estimate the pseudo-continuum from the original X-Shooter spectrum by degrading it to a very low spectral resolution ($\mathrm{R_{\lambda}\,=\,100}$), then subtract this pseudo-continuum from the original section to highlight line features. We add the v.sin(i) as a free-parameter with this fit. We determine the radius analytically with the relation from \cite{2008ApJ...678.1372C}. 
\item Panel (g) shows of zoomed view on a the gravity/metallicity-sensitive lines of K\,I and Na\,I with the best fitting solution. We subtracted the pseudo-continuum in using the same method that in the panel (f). The determination of the radius is analytic too.
\end{itemize}

Generally, the results show that the \texttt{BT-SETTL15}\, models fail to consistently reproduce the pseudo-continuum of the X-Shooter spectra of all objects at all wavelengths (visible and near-infrared) with the same range of physical parameters. The same problem arises when considering a $\mathrm{\chi^{2}}$\, test on the original grid of \texttt{BT-SETTL15}\, spectra. This is highlighted in panels (a)-(d) of all figures, where the best fit solution varies considerably depending on which spectral range is fit. Similar discrepancies were evidenced by \cite{2014AeA...564A..55M} using older releases of the models \citep{2011ASPC..448...91A,2013MSAIS..24..128A, 2015ApJ...802...61L, 2017MNRAS.465..760B, 2012RSPTA.370.2765A}. Therefore, the problem remains in the 2015 release of the models.
	    
The posteriors of \Teff\,are strongly tied to the pseudo-continuum shape. Consequently, they are affected by the choice of the spectral range chosen to estimate the best fit. For instance, the \Teff\, is generally higher from a fit using the J-band compared to fit using H or K-band. The surface gravity is also affected as it remains sensitive to the shape of the VO and H$_{2}$O absorption bands. Notably, the radius, which is linked to the flux dilution factor, seems to be mostly consistent when using three different fitting bands. In addition to the best fitting solution, the \texttt{ForMoSA} code provides the errors on the posterior solutions. The errors on the posterior solutions are tightly connected to the errors bars on the spectra themselves by the likelihood function. Given the high-S/N of the X-Shooter spectra for HIP\,78530\,B, HIP\,77900\,B, USco\,1610-1913\,B, TWA\,26 and TWA\,29, the resulting fitting error bars are about one to two orders of magnitude smaller than the systematics associated with the choice of the spectral fitting window, and the non-perfect match of the \texttt{BT-SETTL15}\, models with our spectra at all observed wavelengths (see Table\,\ref{Tab:Results_FORMOSA1} - \ref{Tab:Results_FORMOSA2}). Our adopted solutions in \Teff, log(g) and R are therefore derived from the average and dispersion of the solutions from the different spectral fitting windows with the extinction value given by \cite{2019arXiv190204116L}. The bolometric luminosity is calculated with the Stefan-Boltzmann law. 
	    
In addition to different spectral fitting window, the \texttt{ForMoSA} code was also applied with the extinction $\mathrm{A_{v}}$ as a free parameter. The resulting posteriors can be directly compared with the ones obtained with the \textit{Gaia} DR2 values at the location of our targets \citep{2019arXiv190204116L}. Surprisingly, with the simple use of an interstellar extinction law correction, we considerably improve the goodness of the fits of all targets (see Figures \ref{Fig:diff_fit_TWA26} - \ref{Fig:diff_fit_HIP77900B} panel (e)). Considering that our targets have no extinction (both sources are located at less than 80\,pc; TWA\,26 has a transition disk around it but the flux excess appears  at $\lambda > 20 \mic$,  \citealt{2008ApJ...681.1584R}; TWA\,29 has no known disk, \citealt{2015AeA...582L...5R}) or precisely known and low interstellar extinction values ($\mathrm{A_{v}}\leq0.13$ for our three Upper Sco companions), the excellent good fit of all objects with artificial $\mathrm{A_{v}}$ values of 1.76-2.58 mag indicates a missing physical component in the models. As all objects are affected, it is very likely that the origin is not circumstellar, but linked to the current state of art of \texttt{BT-SETTL15}\, models as for example the complexity of the dust formation process or the dust opacity \citep{2007MNRAS.380.1285P}.
Systematically testing how this deficiency evolves with lower effective temperature, age (surface gravity) and gravity would be very interesting. 

\begin{figure}
  \centering
  \includegraphics[width=\columnwidth, angle=0]{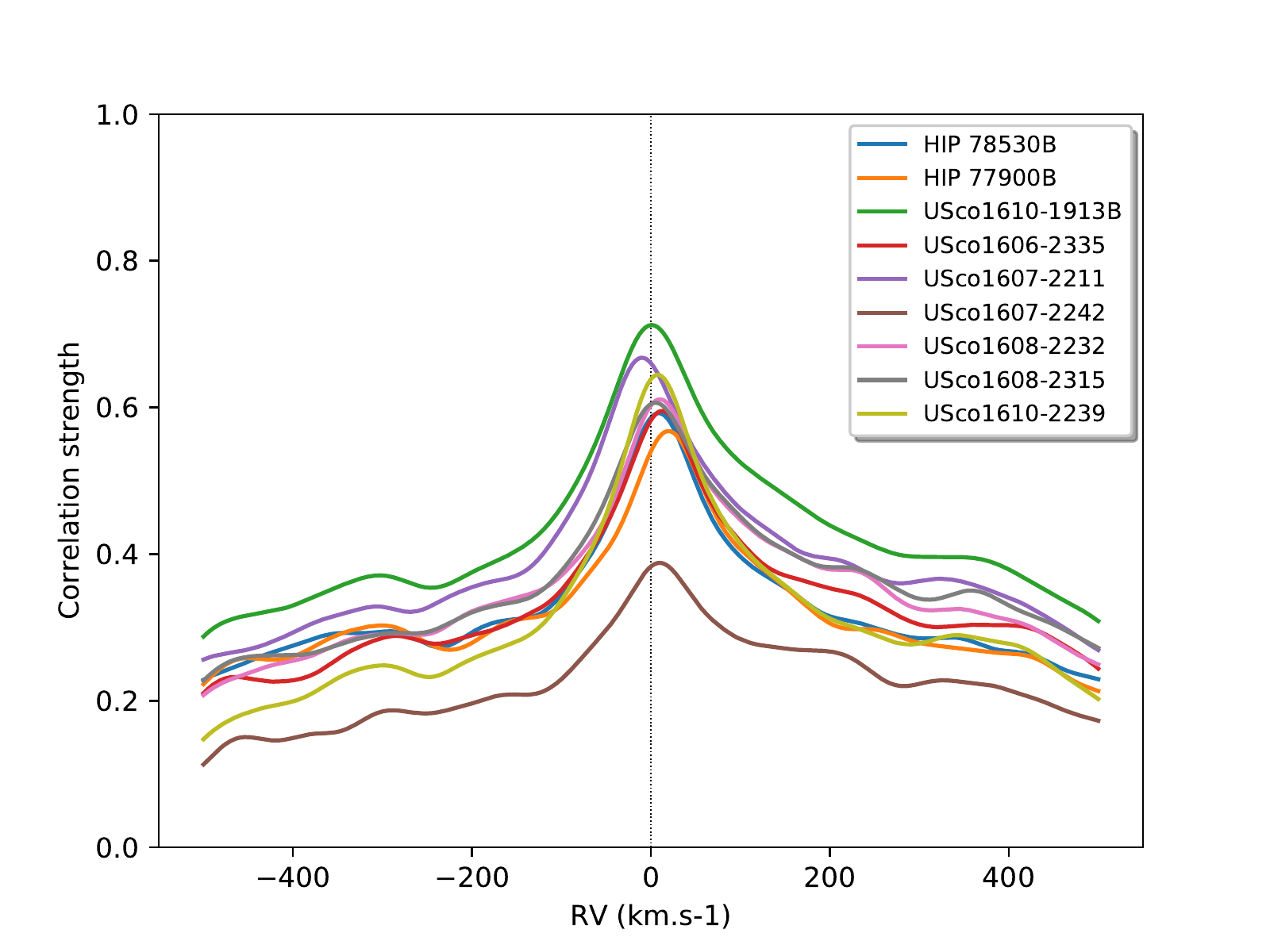}
  \caption{Correlation strength as a function of radial velocity for each of our object. We use the synthetic spectra at \Teff=2400K and log(g)=4.0 dex from \texttt{BT-SETTL15}.}
  \label{Fig:Cross_corr}
\end{figure}

As an additional step to exploit the medium resolution of our X-Shooter spectra, the radial and rotational velocity information (line profile and offset) has been incorporated into \texttt{ForMoSA} as additional fitting parameters. The radial velocity (RV) can be adjusted on the full spectral window. For the adjustment of v.sin(i), we restricted the fitting window to the optical part (0.56-1.00$\,\mu m$) with the continuum subtracted to minimize the computation time. This choice is driven by the higher spectral resolution in optical (R\,=\,5400) and the presence of various absorption lines (NaI, FeH or VO). In Table \ref{Tab:Results_FORMOSA1} - \ref{Tab:Results_FORMOSA2}, one can see that the RV results strongly depends on the choice of the fitting spectral window. We adopted the RV value from the optical fit for the same reasons that for the v.sin(i). The two identified causes of this discrepancy are the slope issues in models and a systematic error in the wavelength calibration of the spectra. We therefore decided to independently compute the RVs using a cross-correlation approach between our observed spectra and a spectral template generated by the \texttt{BT-SETTL15} code at a \Teff\,=\,2400K and a log(g)\,=\,4.0 dex (see Figure \ref{Fig:Cross_corr}). For most cases, there is a good agreement between the cross-correlation results and the \texttt{ForMoSA} ones in the optical that we therefore adopt as final values. For the rotational velocity, the limited X-Shooter spectral resolution does not allow to measure the low v.sin(i) of our targets, but rahter places an upper limit of roughly $\leq$ 50 km/s. This limit is consistent with the 55 km/s limit expected for a resolution of about R = 5400 in the visible arm. Considering the adopted values of \Teff, log(g), R, RV, v.sin(i) and L using the fixed extinction values of \cite{2019arXiv190204116L}, we now detail the results of \texttt{ForMoSA} target by target.

\noindent{\textit{TWA\,26 and TWA\,29\,:}} Both targets are well known and their X-Shooter spectra have been analyzed as a sanity check of the \texttt{ForMoSA} code. They have been classified by \cite{2015AeA...579A..66M} as late-M dwarfs with spectral types M$9\gamma$ and M9.5, respectively. Both are members of the TW Hydrae association, with an age estimate of 8\, Myr, and are located at a distance of 49\,pc and 83\,pc, respectively. They have no extinction observed in their line of sight. Both have been  characterized by \cite{2015ApJ...810..158F} using a complimentary approach to our work. Exploiting the spectral energy distribution from the visible to the near-infrared and the mid-infrared, \cite{2015ApJ...810..158F} derived the bolometric luminosities of these two young brown dwarfs with known distances. 
Using evolutionary model predictions from solar metallicity \texttt{SMHC08} \citep{2008ApJ...689.1327S} and \texttt{DMESTAR} (\citealt{2012ApJ...761...30F}, \citeyear{2013ApJ...779..183F}) isochrones, they derived the predicted radii and masses, and therefore the object effective temperatures using the Stefan-Boltzmann law as well as the surface gravities. The results are reported in Table\,\ref{Tab:comparison_TWA_Fil}, and roughly agree within the error bars with the ones obtained by the \texttt{ForMoSA} forward modelling of the X-Shooter spectra of TWA\,26 and TWA\,29. As the three Upper Sco companions HIP\,78530\,B, HIP\,77900\,B, USco\,1610-1913\,B scan similar range of temperature and surface gravities than TWA\,26 and TWA\,29, this test tend to support the use of \texttt{ForMoSA} to explore their physical properties. These three companions have no known mid-infrared photometry or spectroscopy, therefore a direct determination of the bolometric luminosity as done by \cite{2015ApJ...810..158F} is here not possible
Based on a fit in using the optic wavelength range, we derive a v.sin(i)$\,=\,66\pm2$ km/s for TWA\,29. It is only object for which we are not limited by the resolving power of the instrument.
        

	   

  	    \noindent{\textit{USco\,1610-1913\,B\,:}} For their study of this companion, \cite{2015ApJ...802...61L} used an incorrect value of $\mathrm{A_{v}}$\,=\,1.1 given by \cite{2009ApJ...705.1646C}. Using the method described in \cite{2008ApJS..179..423C}, \cite{2009ApJ...705.1646C} derived this relatively high $\mathrm{A_{v}}$ using the optical (Johnson $B$ and $V$, Tycho $B\mathrm{_{T}}$ and $V\mathrm{_{T}}$) and near-infrared (2MASS $J$, $H$, and $K_s$) photometry and colors, but a wrong spectral type (K7) for USco\,1610-1913. As described in Appendix \ref{sec:AppA}, an updated analysis of the full SED actually indicates a later spectral type M0, leading to an extinction  value of A$\mathrm{_{v}\,=\,0.09\,\pm\,0.01}$ mag, which is more consistent with the value from \cite{2019arXiv190204116L} : $\mathrm{A_{v}}$\,=\,0.13. Using $\mathrm{A_{v}}$\,=\,0.13, we derive with \texttt{ForMoSA} an effective temperature of  \Teff$\,=\,2542^{+68}_{-104}$ K, a surface gravity of $\rm{log(g)}\leq4.17$ dex, both compatible with a young M9 brown dwarf of very low gravity for USco\,1610-1913\,B. However, we find clear non-physical solutions for the radius and the luminosity (for a young late-M dwarf), which confirm the strong over-luminosity of this source, as already pointed out in Section \ref{Sec:Empirical_analysis} and that we will discuss further in Section \ref{Sec:Discussion}. We notice a difference between the radial velocity found from \texttt{ForMoSA} and the one found from the cross-correlation algorithm.

       \noindent{\textit{HIP\,78530\,B\,:}} The extinction maps of \cite{2019arXiv190204116L} gives a value of $\mathrm{A_{v}}$\,=\,0.075 along the line of sight of HIP\,78530, somewhat different from the value of $\mathrm{A_{v}}$\,=\,0.5 used by \cite{2015ApJ...802...61L} and from \cite{2009ApJ...705.1646C}. We adopted the former value in our analysis. \texttt{ForMoSA} finds \Teff$\,=\,2679^{+118}_{-119}$ K, in agreement with the M8 spectral type of the source \citep[this work]{2003ApJ...593.1093L, 2015ApJ...810..158F}. The analysis of the continuum-subtracted spectrum yields a surface gravity estimate compatible with the intermediate gravity class of  the object (Section \ref{Sec:Empirical_analysis}). The luminosity value of \Lum \,=\,$-2.87^{+0.15}_{-0.15}$ is also compatible with expected \Lum \,=\,-2.5 to -3.0 found by \cite{2015ApJ...810..158F} for young M$8\beta$ and M8$\gamma$ BDs. Finally, the effective temperature \Teff\,and surface gravity are consistent with the values from \cite{2011ApJ...730...42L}.
	  
	    \noindent{\textit{HIP\,77900\,B\,:}} With an extinction value of $\mathrm{A_{v}}$\,=\,0.07 \citep{2019arXiv190204116L}, the resulting effective temperature of  \Teff$\,=\,2602^{+117}_{-97}$ K given by \texttt{ForMoSA} is consistent with the later M$9\,\pm\,0.5$ spectral type empirically derived for this companion. The estimated surface gravity of $\rm{log(g)} \,\leq\,4.36\,dex$ is consistent with the intermediate gravity classification of \cite{2013ApJ...772...79A}, and the luminosity value of \Lum \,=\,$-2.89^{+0.15}_{-0.13}$ is  compatible with the values found for young M9$\beta\gamma$ brown dwarfs \citep{2015ApJ...810..158F}.

\begin{table*}[!p]
\caption{Results from fits with \texttt{ForMoSA}.}
\label{Tab:Results_FORMOSA1}
\begin{center}
\begin{center}
\tiny
\renewcommand{\arraystretch}{1.25}
\begin{tabular}{c|ccccccc|c|c}
\hline
\hline
TWA\,26 &	J	&	H  &  K & JHK  & JHK and Av & JHK-cont & OPT-cont & OPT-croscor & Adopted 	\\
\hline
\Teff\,(K)  &  $2412\pm1$  &  $2640\pm1$  &  $2590\pm1$  &  $2114\pm1$  &  $2587\pm1$  &  $2534\pm2$  &  $2522\pm3$  & - & $2547^{+94}_{-136}$\,\,\tnote{a} \\
\hline
log(g)\,(dex)  &  $4.10\pm0.01$  &  \textit{3.5}  &  \textit{3.5}  &  \textit{3.5}  &  \textit{3.5}  &  $3.95\pm0.01$  &  $4.32\pm0.01$  & - &  $\leq\,4.11$\,\,\tnote{b}\\
\hline
R\,(\RJup)  &  $2.61\pm0.01$  &  $2.46\pm0.01$  &  $2.83\pm0.01$  &  $3.72\pm0.01$  &  $3.10\pm0.01$  &  $NA$  &  $NA$    & - & $1.97^{+0.87}_{-0.52}$\,\,\tnote{a} \\
\hline
RV\,(km/s)  &  $-20.0\pm0.2$   &  $10.2\pm0.2$  &  $4.1\pm0.1$   &  $15.1\pm0.1$   &  $5.4\pm0.1$  &  $1.4\pm0.1$    &  $18.2\pm0.2$   & - &  $18.2\pm0.2$ \\
\hline
\Lum &  $-2.68\pm0.01$  &  $-2.57\pm0.01$  &  $-2.48\pm0.01$  &  $-2.60\pm0.01$  &  $-2.41\pm0.01$  &  $NA$  &  $NA$    & - & $-2.83^{+0.38}_{-0.36}$\,\,\tnote{c}\\
\hline
$\mathrm{A_{v}}$\,(mag)  &  \multicolumn{4}{c|}{0.0}  &  \multicolumn{1}{c|}{$2.58\pm0.01$}  &  \multicolumn{2}{c|}{0.0}  & - & 0.0\,\,\tnote{d}  \\
\hline
v.sin(i)\,(km/s)  &  \multicolumn{6}{c|}{-}  &  $\leq\,44$   & - &  $\leq\,44$ \\
\hline
\hline
\multicolumn{8}{c}{}\\
\hline
\hline
TWA\,29 &	J	&	H  &  K & JHK  & JHK and Av & JHK-cont & OPT-cont & OPT-croscor & Adopted 	\\
\hline
\Teff\,(K)   &  $2425\pm2$  &  $2577\pm3$  &  $2564\pm5$  &  $2133\pm1$  &  $2554\pm2$  &  $2519\pm7$  &  $2495\pm7$  & - & $2522^{+58}_{-99}$ \\
\hline
log(g)\,(dex)  &  $4.11\pm0.02$  &  \textit{3.5}  &  $4.31\pm0.02$  &  \textit{3.5}  &  \textit{3.5}  &  $3.99\pm0.02$  &  $4.29\pm0.03$  & - &  $\leq\,4.33$  \\
\hline
R\,(\RJup)  &  $2.08\pm0.01$  &  $2.05\pm0.01$  &  $2.28\pm0.01$  &  $2.92\pm0.01$  &  $2.48\pm0.01$  &  $NA$  &  $NA$    & - & $2.14^{+0.15}_{-0.10}$ \\
\hline
RV\,(km/s)  &  $-21.0\pm1.0$   &  $8.1\pm0.7$   &  $-7.0\pm0.5$   &  $-18.1\pm0.3$  &  $-9.9\pm0.4$   &  $-6.7\pm0.5$   &  $3.2\pm1.0$    & - &  $3.2\pm1.0$  \\
\hline
\Lum  &  $-2.87\pm0.01$  &  $-2.77\pm0.01$  &  $-2.69\pm0.01$  &  $-2.79\pm0.01$  &  $-2.62\pm0.01$  &  $NA$  &  $NA$    & - & $-2.77^{+0.10}_{-0.11}$\\
\hline
$\mathrm{A_{v}\,(mag)}$  &  \multicolumn{4}{c|}{0.0}  &  \multicolumn{1}{c|}{$2.32\pm0.01$}  &  \multicolumn{2}{c|}{0.0} & - & 0.0    \\
\hline
v.sin(i)\,(km/s)  &  \multicolumn{6}{c|}{-}  &  $66\pm2$   & - & $66\pm2$  \\
\hline
\hline
\multicolumn{8}{c}{}\\
\hline
\hline
USco\,1610-1913\,B      &	J	          &	H              &  K             & JHK                               & JHK and Av        & JHK-cont          & OPT-cont          & OPT-croscor   & Adopted 	\\
\hline
\Teff\,(K)              &  $2440\pm2$     &  $2607\pm3$    &  $2580\pm3$    &  \textit{2100}       &  $2652\pm2$       &  $2654\pm8$       &  $2565\pm4$       & -             & $2542^{+68}_{-104}$ \\
\hline  
log(g)\,(dex)           &  $4.16\pm0.01$  &  \textit{3.5}  &  $3.94\pm0.02$  &  \textit{3.5}    &  $3.78\pm0.02$    &  $3.94\pm0.02$    &  $4.23\pm0.02$    & -             &  $\leq\,4.17$  \\
\hline
R\,(\RJup)              &  $3.76\pm0.01$  &  $3.76\pm0.01$ &  $4.10\pm0.01$  &  $5.12\pm0.01$   &  $4.31\pm0.01$    &  $NA$             &  $NA$              & -            & $3.87^{+0.24}_{-0.12}$ \\
\hline
RV\,(km/s)              &  $18.9\pm0.9$   &  $5.3\pm0.6$   &  $-2.4\pm0.4$   &  $13.2\pm0.2$    &  $-2.2\pm0.3$     &  $-0.1\pm0.2$     &  $11.5\pm0.4$     & $0.9$         &  $11.5\pm0.4$ \\
\hline
\Lum                    &  $-2.34\pm0.01$ &  $-2.23\pm0.01$ &  $-2.17\pm0.01$  &  $-2.26\pm0.01$  &  $-2.08\pm0.01$  &  $NA$            &  $NA$             & -             & $-2.25^{+0.10}_{-0.10}$ \\
\hline
$\mathrm{A_{v}\,(mag)}$   &  \multicolumn{4}{c|}{0.13}  &  \multicolumn{1}{c|}{$2.24\pm0.01$}   &  \multicolumn{2}{c|}{0.13}  & - & 0.13   \\
\hline
v.sin(i)\,(km/s)            &  \multicolumn{6}{c|}{-}  &  $\leq\,44$   & - & $\leq\,44$  \\
\hline
\hline
\multicolumn{8}{c}{}\\
\hline
\hline
HIP\,78530\,B &	J	&	H  &  K & JHK  & JHK and Av & JHK-cont & OPT-cont & OPT-croscor & Adopted 	\\
\hline
\Teff\,(K)  &  $2566\pm6$  &  $2791\pm6$  &  $2681\pm8$  &  $2590\pm1$  &  $2774\pm3$  &  $2713\pm13$  &  $2619\pm14$  & - & $2679^{+118}_{-119}$ \\
\hline
log(g)\,(dex)   &  $4.30\pm0.04$  &  $4.00\pm0.03$  &  $3.64\pm0.09$  &  $4.05\pm0.01$  &  $4.06\pm0.01$  &  $4.27\pm0.04$  &  $4.76\pm0.04$  & - &  $\leq\,4.34$  \\
\hline
R\,(\RJup)  &  $1.81\pm0.01$  &  $1.70\pm0.01$  &  $1.98\pm0.01$  &  $1.93\pm0.01$  &  $2.00\pm0.01$  &  $NA$  &  $NA$     & - & $1.83^{+0.16}_{-0.14}$\\
\hline
RV\,(km/s)  &  $-14.0\pm2.2$   &   $-2.0\pm1.3$  &   $-10.7\pm1.2$  &  $-7.8\pm0.7$   &  $-4.1\pm0.8$   & $-4.2\pm0.9$ &  $7.5\pm1.0$   & $8.5$ &  $7.5\pm1.0$ \\
\hline
\Lum   &  $-2.89\pm0.01$  &  $-2.80\pm0.01$  &  $-2.74\pm0.01$  &  $-2.82\pm0.01$  &  $-2.67\pm0.01$  &  $NA$  &  $NA$    & - & $-2.87^{+0.15}_{-0.15}$\\
\hline
$\mathrm{A_{v}\,(mag)}$  &  \multicolumn{4}{c|}{0.075}  &  \multicolumn{1}{c|}{$1.99\pm0.01$}  &  \multicolumn{2}{c|}{0.075}   & - & 0.075  \\
\hline
v.sin(i)\,(km/s)  &  \multicolumn{6}{c|}{-}  &  $\leq\,50$   & - & $\leq\,50$  \\
\hline
\hline
\multicolumn{8}{c}{}\\
\hline
\hline
HIP\,77900\,B &	J	&	H  &  K & JHK  & JHK and Av & JHK-cont & OPT-cont & OPT-croscor & Adopted 	\\
\hline
\Teff\,(K)  &  $2511\pm6$  &  $2713\pm6$  &  $2583\pm13$  &  $2570\pm1$  &  $2713\pm3$  &  $2688\pm17$  &  $2604\pm13$  & - & $2602^{+117}_{-97}$ \\
\hline
log(g)\,(dex)  &  $4.23\pm0.04$  &  $3.58\pm0.05$  &  $4.27\pm0.09$  &  $4.12\pm0.01$  &  $4.03\pm0.02$  &  $4.13\pm0.04$  &  $4.59\pm0.05$  & - &  $\leq\,4.36$  \\
\hline
R\,(\RJup) &  $1.73\pm0.01$  &  $1.65\pm0.01$  &  $1.90\pm0.01$  &  $1.79\pm0.01$  &  $1.88\pm0.01$  &  $NA$  &  $NA$    & - & $1.76^{+0.15}_{-0.12}$ \\
\hline
RV\,(km/s)  &  $-22.7\pm2.3$   &  $-22.7\pm1.3$   &   $-29.5\pm1.6$  &  $-23.9\pm0.8$   &  $-22.8\pm0.9$   &   $-22.9\pm0.8$  &  $19.3\pm1.2$    & $20.1$ &  $19.3\pm1.2$ \\
\hline
\Lum   &  $-2.97\pm0.01$  &  $-2.87\pm0.01$  &  $-2.84\pm0.01$  &  $-2.90\pm0.01$  &  $-2.76\pm0.01$  &  $NA$  &  $NA$   & - & $-2.89^{+0.15}_{-0.13}$ \\
\hline
$\mathrm{A_{v}\,(mag)}$   &  \multicolumn{4}{c|}{0.07}  &  \multicolumn{1}{c|}{$1.76\pm0.01$}  &  \multicolumn{2}{c|}{0.07}   &  - & 0.07 \\
\hline
v.sin(i)\,(km/s)  &  \multicolumn{6}{c|}{-}  &   $\leq\,44$  & - & $\leq\,44$  \\
\hline
\hline
\end{tabular}
\end{center}
\begin{tablenotes}
\item $[$a$]$ Mean and standard deviation between $J$, $H$ and $K$ bands.
\item $[$b$]$ High value between $J$, $H$, $K$ and $JHK$-cont bands.
\item $[$c$]$ Stefan$-$Boltzmann law in using \Teff\,and radius ranges.
\item $[$d$]$ From \cite{2019arXiv190204116L}’s maps (derived from \cite{2018MNRAS.477L..50G}). 
\end{tablenotes}
\end{center}
\end{table*}

\begin{table*}[!p]
\caption{Results from fits with \texttt{ForMoSA}.}
\label{Tab:Results_FORMOSA2}
\begin{center}
\begin{center}
\tiny
\renewcommand{\arraystretch}{1.25}
\begin{tabular}{c|ccccccc|c|c}
\hline
\hline
USco\,1606-2335             &	J	            &	H               &  K             & JHK              & JHK and Av     & JHK-cont      & OPT-cont      & OPT-croscor & Adopted 	\\
\hline
\Teff\,(K)                  &  $2390\pm12$      & $2620\pm13$       & $2547\pm18$    & $2350\pm5$       & $2650\pm7$     & $2574\pm38$   & $2414\pm45$   &  -  & $2519^{+114}_{-141}$  \\ 
\hline
log(g)\,(dex)               &  $4.13\pm0.08$    &  \textit{3.5}     & \textit{3.5}   & \textit{3.5}     & $3.89\pm0.06$  & $4.28\pm0.08$ & \textit{5.0}  &  - & $\leq\,4.36$  \\ 
\hline
R\,(\RJup)                  &  $1.41\pm0.01$    &  $1.37\pm0.01$    & $1.59\pm0.02$  & $1.65\pm0.01$    & $1.66\pm0.01$  & $NA$          & $NA$          &  -    & $1.46^{+0.15}_{-0.10}$  \\ 
\hline
RV\,(km/s)                  &  $-11.7\pm4.9$    &  $-0.9\pm1.6$     & $-10.0\pm2.8$  & $-10.1\pm1.2$    & $-3.4\pm1.5$   & $-4.3\pm1.6$  & $13.6\pm3.6$  &  $13.1$ & $10.1\pm4.1$  \\ 
\hline
\Lum                        &  $-3.23\pm0.01$   &  $-3.09\pm0.01$   & $-3.01\pm0.01$ & $-3.12\pm0.01$   & $-2.91\pm0.01$ & $NA$          & $NA$          &  -    & $-3.11^{+0.16}_{-0.16}$  \\ 
\hline
$\mathrm{A_{v}}$\,(mag)     &  \multicolumn{4}{c|}{0.0}  &  \multicolumn{1}{c|}{$3.06\pm0.03$}  &  \multicolumn{2}{c|}{0.0}  &  - & 0.0\,\,\tnote{d}  \\
\hline
v.sin(i)\,(km/s)  &  \multicolumn{6}{c|}{-}  &  $\leq\,45$   &  - & $\leq\,45$  \\
\hline
\hline
\multicolumn{8}{c}{}\\
\hline
\hline
USco\,1607-2211             &	J	            &	H               &  K              & JHK             & JHK and Av     & JHK-cont         & OPT-cont     & OPT-croscor & Adopted 	\\
\hline
\Teff\,(K)                  &  $2445\pm5$       & $2616\pm6$        & $2609\pm5$      & $2384\pm1$      & $2621\pm2$     & $2674\pm13$       & $2532\pm15$  &  - & $2557^{+65}_{-117}$  \\ 
\hline
log(g)\,(dex)               &  $3.92\pm0.07$    &  \textit{3.5}     & $3.79\pm0.06$   & \textit{3.5}    & $4.20\pm0.01$  & $4.02\pm0.03$    & $4.25\pm0.06$  &  - & $\leq\,4.05$  \\ 
\hline
R\,(\RJup)                  &  $1.92\pm0.01$    &  $1.87\pm0.01$    & $1.96\pm0.01$   & $2.18\pm0.01$   & $2.07\pm0.01$  & $NA$             & $NA$  &  - & $1.92^{+0.05}_{-0.06}$  \\ 
\hline
RV\,(km/s)                  &  $-31.0\pm1.8$    &  $-28.2\pm1.1$    & $-0.5\pm0.6$    & $-14.2\pm0.4$   & $-8.3\pm0.5$   & $-9.1\pm0.5$     & $-10.1\pm1.3$  &  $-10.2$ & $-11.1\pm1.4$  \\ 
\hline
\Lum                        &  $-2.92\pm0.01$   &  $-2.83\pm0.01$   & $-2.79\pm0.01$  & $-2.85\pm0.01$  & $-2.74\pm0.01$ & $NA$             & $NA$  &  - & $-2.84^{+0.07}_{-0.11}$  \\ 
\hline
$\mathrm{A_{v}\,(mag)}$     &  \multicolumn{4}{c|}{0.0}  &  \multicolumn{1}{c|}{$1.64\pm0.01$}  &  \multicolumn{2}{c|}{0.0} &  - & 0.0    \\
\hline
v.sin(i)\,(km/s)  &  \multicolumn{6}{c|}{-}  &  $\leq\,45$   &  - & $\leq\,45$  \\
\hline
\hline
\multicolumn{8}{c}{}\\
\hline
\hline
USco\,1607-2242             &	J	            &	H               &  K              & JHK             & JHK and Av & JHK-cont & OPT-cont & OPT-croscor & Adopted 	\\
\hline
\Teff\,(K)                  &  $2265\pm14$       & $2512\pm13$       & $2432\pm23$    & \textit{2100}   & $2579\pm8$  & $2432\pm71$  & $2380\pm102$  &  - & $2403^{+122}_{-152}$  \\ 
\hline
log(g)\,(dex)               &  $3.84\pm0.07$    &  \textit{3.5}     & \textit{3.5}    & \textit{3.5}    & \textit{3.5}  & $3.96\pm0.14$  & $4.01\pm0.27$  &  - & $\leq\,4.10$  \\ 
\hline
R\,(\RJup)                  &  $1.09\pm0.02$    &  $1.06\pm0.01$    & $1.24\pm0.02$   & $1.47\pm0.01$   & $1.31\pm0.01$  & $NA$  & $NA$  &  - & $1.13^{+0.13}_{-0.08}$  \\ 
\hline
RV\,(km/s)                  &  $-19.0\pm6.1$    &  $-0.1\pm1.2$     & $-5.3\pm3.3$    & $-9.9\pm1.2$    & $-1.1\pm1.4$  & $-1.1\pm1.5$  & $6.7\pm11.6$  &  $9.6$ & $5.7\pm10.8$  \\ 
\hline
\Lum                        &  $-3.55\pm0.01$   &  $-3.39\pm0.01$   & $-3.31\pm0.01$  & $-3.42\pm0.01$  & $-3.16\pm0.01$  & $NA$  & $NA$  &  - & $-3.41^{+0.18}_{-0.18}$  \\ 
\hline
$\mathrm{A_{v}\,(mag)}$     &  \multicolumn{4}{c|}{0.0}  &  \multicolumn{1}{c|}{$3.82\pm0.04$}  &  \multicolumn{2}{c|}{0.0}  &  - & 0.0   \\
\hline
v.sin(i)\,(km/s)  &  \multicolumn{6}{c|}{-}  &  $\leq\,45$   &  - & $\leq\,45$  \\
\hline
\hline
\multicolumn{8}{c}{}\\
\hline
\hline
USco\,1608-2232             &	J	            &	H               &  K                & JHK            & JHK and Av       & JHK-cont & OPT-cont & OPT-croscor & Adopted 	\\
\hline
\Teff\,(K)                  &  $2317\pm7$       & $2482\pm8$        & $2427\pm12$       & \textit{2100}  & $2555\pm5$       & $2510\pm26$  & $2363\pm50$  &  - & $2409^{+81}_{-99}$  \\ 
\hline
log(g)\,(dex)               &  $3.85\pm0.05$    &  \textit{3.5}     & \textit{3.5}      & \textit{3.5}   & \textit{3.5}     & $4.01\pm0.08$  & \textit{5.0}  &  - & $\leq\,4.09$  \\ 
\hline
R\,(\RJup)                  &  $1.52\pm0.01$    &  $1.56\pm0.01$    & $1.81\pm0.01$     & $2.13\pm0.01$  & $1.91\pm0.01$    & $NA$  & $NA$  &  - & $1.63^{+0.19}_{-0.12}$  \\ 
\hline
RV\,(km/s)                  &  $-14.1\pm3.6$     &  $-0.5\pm1.2$     & $-3.5\pm1.9$      & $-10.5\pm0.8$  & $-1.4\pm1.1$     & $-1.5\pm1.2$  & $11.1\pm4.0$  &  $9.7$ & $11.0\pm4.5$  \\ 
\hline 
\Lum                        &  $-3.22\pm0.01$   &  $-3.08\pm0.01$   & $-2.99\pm0.01$    & $-3.10\pm0.01$ & $-2.85\pm0.01$   & $NA$  & $NA$  &  - & $-3.09^{+0.15}_{-0.14}$  \\
\hline
$\mathrm{A_{v}\,(mag)}$     &  \multicolumn{4}{c|}{0.0}  &  \multicolumn{1}{c|}{$3.65\pm0.02$}  &  \multicolumn{2}{c|}{0.0}   &  - & 0.0  \\
\hline
v.sin(i)\,(km/s)  &  \multicolumn{6}{c|}{-}  &  $\leq\,47$   &  - & $\leq\,47$  \\
\hline
\hline
\multicolumn{8}{c}{}\\
\hline
\hline
USco\,1608-2315             &	J               &	H               &  K                & JHK            & JHK and Av       & JHK-cont & OPT-cont & OPT-croscor & Adopted 	\\
\hline
\Teff\,(K)                  &  $2344\pm4$       & $2555\pm5$        & $2561\pm7$        & \textit{2100}     & $2659\pm3$       & $2664\pm15$  & $2494\pm27$  &  - & $2487^{+81}_{-147}$  \\ 
\hline
log(g)\,(dex)               &  $3.80\pm0.03$    &  \textit{3.5}     & \textit{3.5}      & \textit{3.5}   & $3.66\pm0.02$    & $4.13\pm0.03$  & $4.27\pm0.10$  &  - & $\leq\,4.16$  \\ 
\hline
R\,(\RJup)                  &  $1.96\pm0.01$    &  $1.92\pm0.01$    & $2.11\pm0.01$     & $2.78\pm0.01$  & $2.26\pm0.01$    & $NA$  & $NA$  &  - & $2.00^{+0.12}_{-0.09}$  \\ 
\hline
RV\,(km/s)                  &  $-15.4\pm2.1$    &  $-0.1\pm0.5$     & $-1.8\pm1.0$      & $-13.8\pm0.4$   & $-0.2\pm0.3$     & $-0.1\pm0.2$  & $10.6\pm2.3$  &  $5.3$ & $10.4\pm2.5$  \\ 
\hline
\Lum                        &  $-2.98\pm0.01$   &  $-2.84\pm0.01$   & $-2.76\pm0.01$    & $-2.86\pm0.01$ & $-2.63\pm0.01$   & $NA$  & $NA$  &  - & $-2.86^{+0.11}_{-0.15}$  \\ 
\hline
$\mathrm{A_{v}\,(mag)}$     &  \multicolumn{4}{c|}{0.0}  &  \multicolumn{1}{c|}{$3.28\pm0.01$}  &  \multicolumn{2}{c|}{0.0}   &  - & 0.0  \\
\hline
v.sin(i)\,(km/s)  &  \multicolumn{6}{c|}{-}  &  $\leq\,45$   &  - & $\leq\,45$  \\
\hline
\hline
\multicolumn{8}{c}{}\\
\hline
\hline
USco\,1610-2239             &	J               &	H               &  K                & JHK            & JHK and Av       & JHK-cont & OPT-cont & OPT-croscor & Adopted 	\\
\hline
\Teff\,(K)                  &  $2397\pm6$       & $2596\pm5$        & $2503\pm7$        & $2343\pm2$     & $2633\pm3$     & $2612\pm16$  & $2514\pm19$  &  - & $2499^{+102}_{-108}$  \\ 
\hline
log(g)\,(dex)               &  $3.87\pm0.05$    &  \textit{3.5}     & \textit{3.5}      & \textit{3.5}   & \textit{3.5}    & $3.97\pm0.04$  & $4.31\pm0.08$  &  - & $\leq\,4.01$  \\ 
\hline
R\,(\RJup)                  &  $1.87\pm0.01$    &  $1.85\pm0.01$    & $2.15\pm0.01$     & $2.21\pm0.01$  & $2.21\pm0.01$    & $NA$  & $NA$  &  - & $1.98^{+0.24}_{-0.14}$  \\ 
\hline
RV\,(km/s)                  &  $-13.0\pm2.1$    &  $-0.1\pm0.4$     & $-3.4\pm1.1$      & $-7.6\pm0.5$   & $-0.9\pm0.6$     & $-0.5\pm0.6$  & $7.7\pm1.7$  &  $7.9$ & $11.3\pm1.8$  \\ 
\hline
\Lum                        &  $-2.98\pm0.01$   &  $-2.85\pm0.01$   & $-2.78\pm0.01$    & $-2.87\pm0.01$ & $-2.67\pm0.01$   & $NA$  & $NA$  &  - & $-2.86^{+0.17}_{-0.14}$  \\ 
\hline
$\mathrm{A_{v}\,(mag)}$     &  \multicolumn{4}{c|}{0.0}  &  \multicolumn{1}{c|}{$2.91\pm0.01$}  &  \multicolumn{2}{c|}{0.0}   &  - & 0.0  \\
\hline
v.sin(i)\,(km/s)  &  \multicolumn{6}{c|}{-}  &  $\leq\,45$   &  - &  $\leq\,45$ \\
\hline
\hline
\end{tabular}
\end{center}
\end{center}
\end{table*}

\section{Emission line properties}
\label{sec:Emission line properties}
\begin{figure*}[t]
  \centering
  \begin{tabular}{cccc}
    \includegraphics[width=4.cm, angle=0]{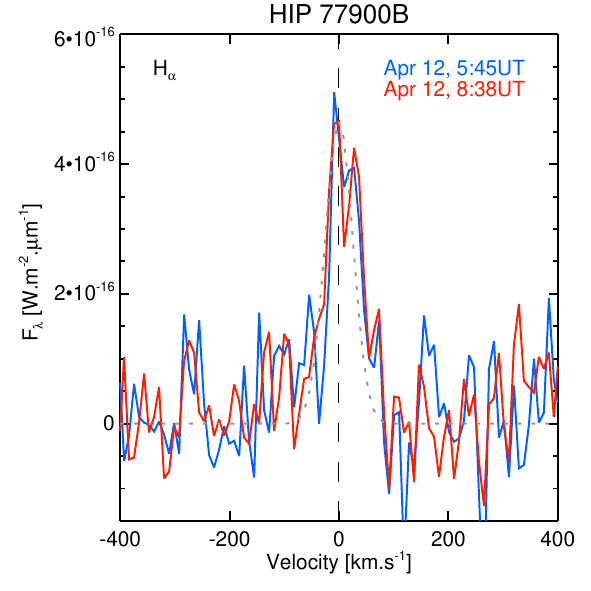} & 
    \includegraphics[width=4.cm, angle=0]{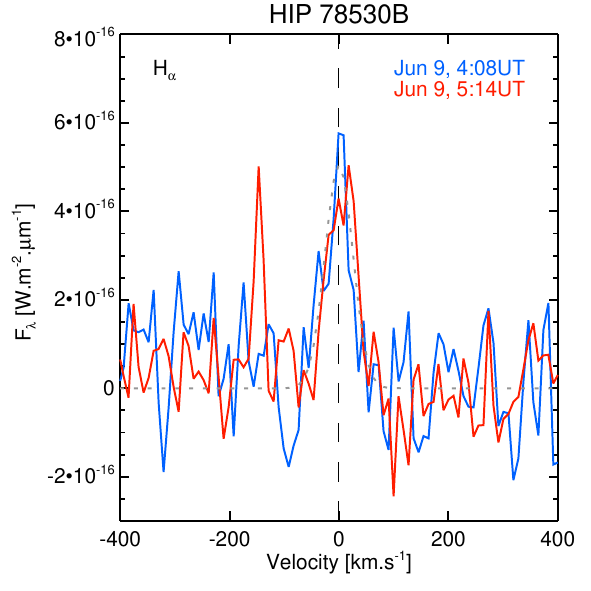} &  
 \includegraphics[width=4.cm, angle=0]{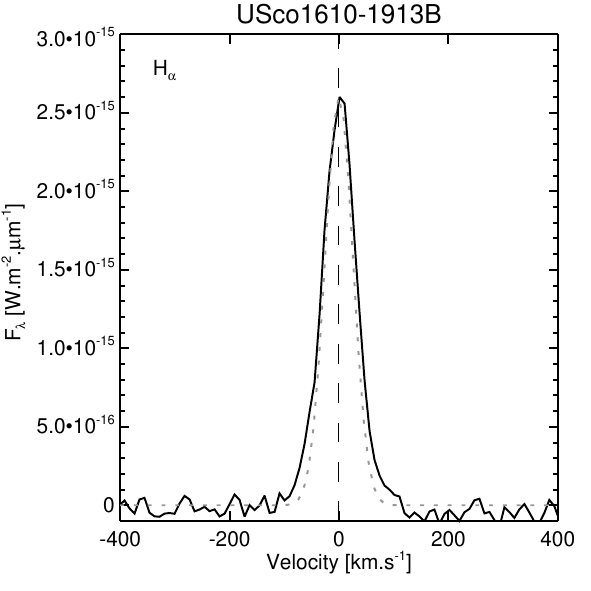} \\ 

    \includegraphics[width=4.cm, angle=0]{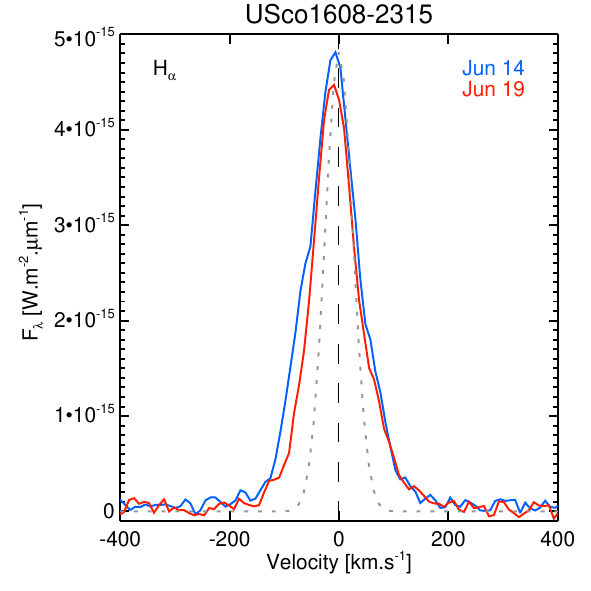} &       \includegraphics[width=4.cm, angle=0]{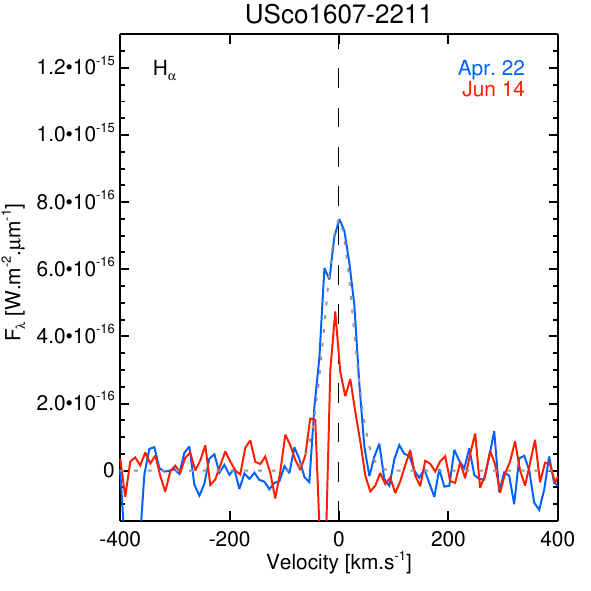} & 
\includegraphics[width=4.cm, angle=0]{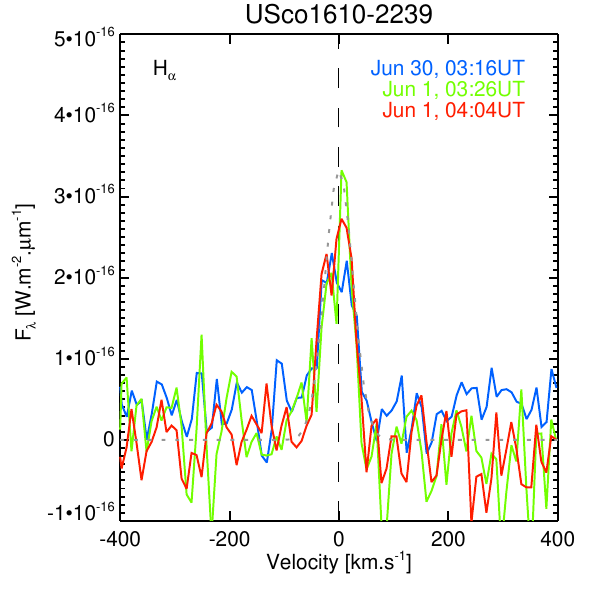} & 
 \includegraphics[width=4.cm, angle=0]{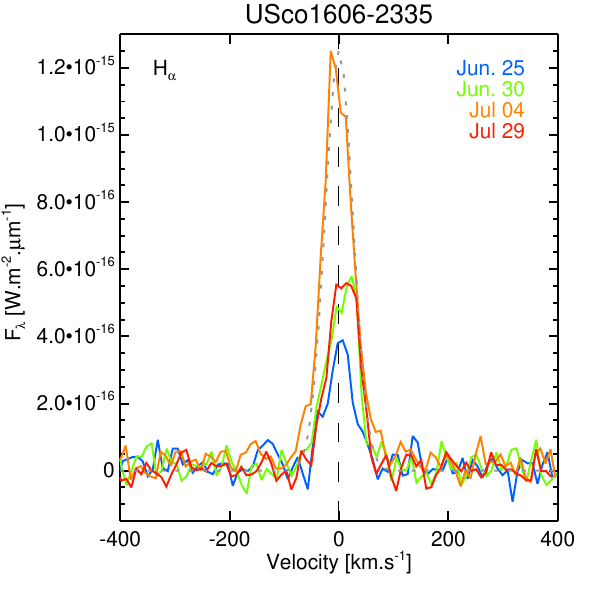}   \\ 
  \end{tabular} 
  \caption{\Ha~line profiles for six of the targets with the strongest line emission. The dotted grey line corresponds to an estimate of the line spread function.}
  \label{Fig:Lineprofile}
\end{figure*}

Our new high S/N medium-resolution optical spectra allow us to identify the faint \Ha\, (656.28 nm) emission lines of the three companions and of four out of the six free-floating objects of our sample. Figure \ref{Fig:Lineprofile} shows the line profiles at each epoch. We checked for the emission line detection in the traces to ensure it was not produced by an uncorrected bad pixel (see Section \ref{Sec:Observations}). \Ha\, lines are commonly attributed to magnetospheric accretion (\citealt{2013AeA...551A.107M, 2015AeA...579A..66M, 2004AeA...424..603N}) or chromospheric activity \citep[e.g.,][]{2003ApJ...582.1109W}. The three objects (USco\,1610-1913\,B, USco\,1606-2335, and USco\,1608-2315) showing the strongest \Ha\, lines also display other Balmer lines, \Hb\, (486.1 nm) and \Hg\, (434.0 nm), in their spectra. We  tentatively identify Ca\,II-H (393.4 nm) and Ca\,II-K (396.9 nm) emission lines in the spectrum of USco\,1610-1913\,B as well. The Ca II lines can also be related to accretion or activity \citep{2008ApJ...681..594H}. We report the apparent fluxes of these emission lines and 10\% width of \Ha\, line in Table \ref{Tab:lines}. The noisy spectral continua of our objects prevented us from obtaining robust determination of the continuum and the equivalent widths. Both origins (accretion and activity) for our targets are investigated below.

 \begin{figure}[t]
  \centering
  \includegraphics[width=\columnwidth, angle=0]{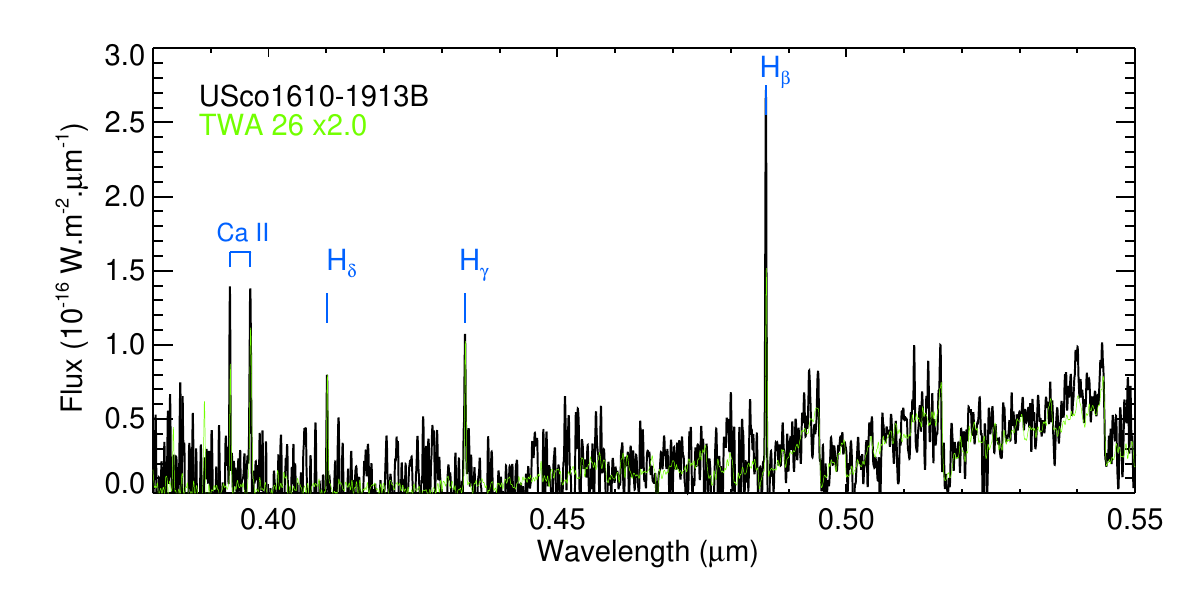}
  \caption{Comparison of the UVB spectrum of USco\,1610-1913\,B (\textit{black}) and TWA\,26 (green) re-normalized to match the spectral continuum of the companion. Emission lines identified in the spectra are reported.}
  \label{Fig:CompUVB}
\end{figure}

\subsection{Accretion rate determination}

The accretion rates of each object are reported in Table \ref{Tab:accretion_rate}. They were derived following the relation of  \cite{2017AeA...600A..20A}:

\begin{equation}
\dot{M}_{acc}\,=\,\left(1-\frac{R_{obj}}{R_{in}}\right)^{-1}\frac{L_{acc}R_{obj}}{GM_{obj}}\approx 1.25\frac{L_{acc}R_{obj}}{GM_{obj}}
\end{equation}

where $\mathrm{L_{acc}}$ is the accretion luminosity, $\mathrm{R_{in}}$ is the inner disk radius, $\mathrm{R_{obj}}$ and $\mathrm{M_{obj}}$ are the radius and mass of the objects, respectively. This is assuming $\mathrm{R_{obj}}$/$\mathrm{R_{in}}$ = 0.2 \citep{2017AeA...600A..20A} which holds in our mass range and whatever the object configuration (companion or isolated). The accretion luminosities were estimated beforehand from the line fluxes of the  Ca\,II lines and the Balmer lines using the relationships given in \cite{2012AeA...548A..56R} and \cite{2017AeA...600A..20A}, respectively. Radii and masses were in addition inferred from 
the evolutionary model predictions as indicated in Table \ref{Tab:param_compare}. 

Alternatively, we derived accretion rates based on 10\% \Ha\, width following the relations of \cite{2004AeA...424..603N}: 
\begin{equation}
log(\dot{M}_{acc})\,=\,-12.89(\pm0.3) + 9.7(\pm0.7) \times 10^{-3} H_{\alpha}10\%    
\end{equation}

The 10\% \Ha\, accretion rates are also reported in Table \ref{Tab:accretion_rate}. They tend to be higher than the ones inferred from the Balmer and Ca II\,line fluxes. A similar divergence between the different accretion proxies is found for the young companion SR 12C \citep{2018MNRAS.475.2994S}.  The accretion rates of USco\,1610-1913\,B are compatible with those of the 1-5 Myr late-M sub-stellar companions FW\,Tau\,b \citep{2014ApJ...784...65B},  DH\,Tau\,b \citep{2014ApJ...783L..17Z}, and GQ\,Lup\,b \citep{2017ApJ...836..223W}, which had the lowest recorded accretion rates among known accreting companions. The accretion rates of HIP\,78530\,B and HIP\,77900\,B are found to be an order of magnitude lower, this indicates these objects as non-accretor. 

\subsection{Chromospheric activity}

The weakness of the lines suggests the emission arises mostly from chromospheric activity. \cite{2013AeA...551A.107M} provides the $\mathrm{L_{acc, noise}/L_{bol}}$ values expected from chromospheric activity for young stars and brown-dwarfs, down to the M9 spectral type:
\begin{equation}
    log(L_{acc, noise}/L_{bol})\,=\,(6.17 \pm 0.53) \times log(\Teff) - (24.54 \pm 1.88)
\end{equation}

All our objects but USco\,1608-2315 have $\mathrm{log(L_{acc}/L_{bol})}$ values below or comparable to the activity threshold  $\mathrm{log(L_{acc, noise}/L_{bol})}$ corresponding to their \Teff\, (estimated from \texttt{ForMoSA}), thus indicating that the lines are indeed likely related to chromospheric activity. \cite{2009ApJ...696.1589H} also identify a \Ha\, line on USco\,1607-2211 of similar intensity and reach the same conclusion. We note that this criterion applied to the \Ha\,line of GQ\,Lup\,b as observed on 2015, April 2016 \citep{2017ApJ...836..223W} indicates that the line mostly arises from chromospheric activity in spite of accretion at this epoch\footnote{We considered the \Teff\, from \cite{2007AeA...463..309S} and \cite{2014ApJ...783L..17Z} and the luminosity from \cite{2014ApJ...783L..17Z} to compute the $\mathrm{log(L_{acc, noise}/L_{bol})}$\,=\,-3.8 to -3.48 value and compare it to  $\mathrm{log(L_{acc}/L_{bol})}\,=\,-3.9\mathrm{\: to \:} -2.9$ of the object.}. The 10\% width of the \Ha\,line may also discriminate between the accretion (large values) and chromospheric activity (small values). A threshold of 200 km.s$^{-1}$ is generally adopted for brown-dwarfs \citep{2003AJ....126.1515J}. All of our objects but USco\,1608-2315 have an 10\% \Ha\, below the 200 km.s$^{-1}$ limit, thus supporting line production by chromospheric activity. Therefore, USco\,1608-2315 is likely the only true accreting object in our sample. The source also has a tentative disk excess (Section \ref{sec:Targets}). Thus, we add USco\,1608-2315 to the short list of free-floating accretors with masses below 30 \MJup\, \citep[][]{2009ApJ...696.1589H, 2012AeA...547A..80B, 2013AeA...558L...7J, 2014AeA...561A...2A,  2016ApJ...832...50B, 2018MNRAS.473.2020L}. 

To further support these conclusions for a few specific objects, Figure \ref{Fig:CompUVB} compares the UVB spectra of USco\,1610-1913\,B and TWA\,26 scaled to the distance of USco\,1610-1913\,B. The flux from TWA\,26 had to be multiplied by an additional factor of two to match USco\,1610-1913\,B's flux level (see Section \ref{subsec:overlum}). The pseudo-continuum of USco\,1610-1913\,B is well reproduced by a template with emission lines due to chromospheric activity \citep{2013AeA...551A.107M}. The \Hg\, and \Hd\ lines have similar strengths but the \Hb\, line appears more luminous in the companion spectrum. 

Figure \ref{Fig:Lineprofile} shows that the two free-floating objects with the strongest lines (USco\,1608-2315 and USco\,1606-2335) display significant short-term variability. The lines are not detected in the optical spectra of these objects obtained in May 2007 \citep{2011AeA...527A..24L}. The \Ha\, lines of USco\,1607-2211 and USco\,1610-2239 do not show obvious variability in our data, but are not detected in May 2007 \citep{2011AeA...527A..24L} as well. It is possible that the non-detections from 2007 arise from a degraded sensitivity of the low resolution data from \cite{2011AeA...527A..24L}. Active M-dwarfs are known to display such variability from chromospheric activity \citep[e.g.,][]{2002AJ....123.3356G}, which strengthens our conclusions on the line origins. The variability of USco\,1608-2315 could also stem from variation of the accretion rate (Aguayo et al., in prep; Bonnefoy et al., in prep.).

\begin{table*}
 \centering
\renewcommand{\arraystretch}{1.3}
  \caption{Apparent fluxes of the emission lines for the companions and free-floating brown dwarfs from our original sample. We also report the 10\% width  of the \Ha\,  line. The upper limits on the line flux correspond to the 1$\sigma$ noise level of the continuum.}
\tiny
  \begin{tabular}{cc|ccccc|c}
  \hline   
	\hline
Object			& Date	&		Ca II-H	&	Ca II-K	&	\Hg	&	\Hb	&	\Ha	&	10\%~\Ha  	  \\
				&		&	($\mathrm{10^{-19} W.m^{-2}}$)		&		($\mathrm{10^{-19} W.m^{-2}}$)		&		($\mathrm{10^{-19} W.m^{-2}}$)	&	($\mathrm{10^{-19} W.m^{-2}}$)		&		($\mathrm{10^{-19} W.m^{-2}}$)	&  ($\mathrm{km.s^{-1}}$) \\
\hline
USco\,1610-1913\,B & 2014-04-02 & $0.27\pm0.11$	&	 $0.23\pm0.15$	&	 $0.23\pm0.18$ &	$0.56\pm0.21$	&	$4.86\pm0.26$&	145  \\
HIP\,77900\,B	& 2014-04-12 05:45UT & $ <0.31$	& $<0.22$	&	$<0.45$	&	$<0.46$	&	$0.40\pm0.15$  & \dots   \\
HIP\,77900\,B	& 2014-04-12 08:38UT & $ <0.20$	& $<0.23$	&	$<0.24$	&	$<0.37$	&	$0.58\pm0.15$  & \dots   \\
HIP\,78530\,B	& 2014-06-09 04:08UT & $<0.95$		& $<0.99$		&	$<1.28$		&	$<2.61$	&	$0.68\pm0.36$  & \dots  \\
HIP\,78530\,B	& 2014-06-09 05:14UT & $<0.47$		& $<0.72$		&	$<0.79$		&	$<2.19$	&	$0.61\pm0.29$  & \dots  \\
\hline
USco\,1607-2242 & all epochs &  $<0.06$ & $<0.10$&	$<0.07$	&	$<0.09$	&	$<0.10$  &  \dots   \\
USco\,1608-2232 & all epochs	& $<0.04$	& $<0.04$	&	$<0.03$	&	$<0.10$	&	$<0.14$	  &  \dots    \\
USco\,1606-2335 & 2014-06-25 &	$<0.15$	& $<0.13$	&	$ 0.14\pm0.14$ &	$0.23\pm0.17$	&	$0.53\pm0.19$  & \dots   \\
USco\,1606-2335  & 2014-06-30 &	$<0.09$	& $<0.32$	&	$ 0.22\pm0.19$ &	$0.18\pm0.11$	&	$0.72\pm0.17$  & 169 \\ 
USco\,1606-2335 & 2014-07-04 &	$<0.07$	& $<0.09$	&	$ 0.24\pm0.12$ &	$0.39\pm0.11$	&	$1.58\pm0.15$  & 136  \\
USco\,1606-2335 & 2014-07-29 &	$<0.08$	& $<0.09$	&	$ 0.18\pm0.10$ &	$0.22\pm0.08$	&	$0.85\pm0.15$  & 105:  \\
USco\,1610-2239   &  2014-06-30 & $<0.10$	& $<0.10$ 	&	$<0.15$	&	$<0.13$	&	$0.28\pm0.09$  &  \dots   \\
USco\,1610-2239   & 2014-07-01 03:26UT & $<0.16$	& $<0.93$ 	&	$<0.19$	&	$<0.19$	&	$0.41\pm0.16$  &  \dots   \\
USco\,1610-2239   & 2014-07-01 04:04UT & $<0.07$	& $<0.11$ 	&	$<0.13$	&	$<0.12$	&	$0.39\pm0.11$  &  \dots   \\
USco\,1608-2315 & 2014-06-15 & $<0.11$	& $<0.29$	&	$1.07\pm0.23$	&	$1.85\pm0.25$	&	$11.99\pm0.38$  & 220  \\
USco\,1608-2315 & 2014-06-20 & $<0.11$	& $<0.25$	&	$0.83\pm0.14$	&	$1.07\pm0.14$	&	$7.65\pm0.30$  & 209  \\
USco\,1607-2211 & 2014-04-22	&	$<0.17$	& $<0.15$	&	$<0.20$	&	$<0.23$	& $0.99\pm0.18$  & 160  \\
USco\,1607-2211 & 2014-06-14	&	$<0.11$	& $<0.11$	&	$<0.17$	&	$<0.23$	& \dots  & \dots    \\
\hline
\hline
 \end{tabular}
\label{Tab:lines}
\end{table*}

\begin{table*}
\centering
\renewcommand{\arraystretch}{1.3}
 \caption{Accretion luminosities computed from the emission lines}
\tiny
 \begin{tabular}{ccccccc|c|c}
 \hline   
	\hline
Object			& Date	&	Ca II-H	&	Ca II-K	&	\Hg	&	\Hb	&	\Ha	& 	 Balmer & \\
				&	    & \Lumacc	& \Lumacc   & \Lumacc & \Lumacc &	\Lumacc	 & 	\Lumacc & $\mathrm{log(L_{acc, noise}/L_{\odot})}$ \\
\hline
USco\,1610-1913\,B & 2014-04-02 & $-5.57\pm0.21$ & $-5.56\pm0.37$ & $-6.00\pm0.70$ & $-5.89\pm0.14$ & $-5.61\pm0.05$ & $-5.70 \pm 0.30$ & $-5.65$\\
HIP\,77900\,B	& 2014-04-12 05:45UT & $\leq-5.52$ & $\leq-5.65$ & $\leq-5.67$ & $\leq-6.06$ & $-6.79\pm0.05$ & $-6.79\pm0.05$  & $-6.06$ \\
HIP\,77900\,B	& 2014-04-12 08:38UT & $\leq-5.73$ & $\leq-5.63$ & $\leq-5.98$ & $\leq-6.05$ & $-6.61\pm0.02$ & $-6.61\pm0.02$ & $ -6.07$ \\
HIP\,78530\,B	& 2014-06-09 04:08UT & $\leq-5.04$ & $\leq-4.95$ & $\leq-5.22$ & $\leq-5.18$ & $-6.62\pm0.18$ &  $-6.62\pm0.18$ & $-6.07$ \\
HIP\,78530\,B	& 2014-06-09 05:14UT & $\leq-5.36$ & $\leq-5.09$ & $\leq-5.45$ & $\leq-5.26$ & $-6.67\pm0.13$ & $-6.67\pm0.13$  & $-6.07$ \\
\hline
USco\,1607-2242 & all epochs &  $\leq-6.26$ & $\leq-5.93$ & $\leq-6.56$ & $\leq-6.79$ & $\leq-7.51$ & \dots & \dots   \\
USco\,1608-2232 & all epochs	& $\leq-6.45$ & $\leq-6.34$ & $\leq-6.97$ & $\leq-6.74$ & $\leq-7.34$ & \dots & \dots \\
USco\,1606-2335 & 2014-06-25  & $\leq-5.84$ & $\leq-5.81$ & $-6.20\pm1.34$ & $-6.33\pm0.63$ & $-6.69\pm0.16$ & $-6.58 \pm 0.71$  & $-6.60$ \\
USco\,1606-2335  & 2014-06-30 & $\leq-6.07$ & $\leq-5.41$ & $-6.01\pm1.01$ & $-6.45\pm0.44$ & $-6.54\pm0.09$ & $-6.49 \pm 0.51$  & $-6.60$ \\ 
USco\,1606-2335 & 2014-07-04  & $\leq-6.19$ & $\leq-5.98$ & $-5.97\pm0.40$ & $-6.06\pm0.16$ & $-6.15\pm0.03$ & $-6.13 \pm 0.20$ & $-6.60$ \\
USco\,1606-2335 & 2014-07-29  & $\leq-6.13$ & $\leq-5.98$ & $-6.11\pm0.45$ & $-6.35\pm0.20$ & $-6.46\pm0.06$ & $-6.40 \pm 0.24$ & $-6.60$\\
USco\,1610-2239   &  2014-06-30 & $\leq-6.13$ & $\leq-6.04$ & $\leq-6.31$ & $\leq-6.73$ & $-7.12\pm0.09$ & $-7.12\pm0.09$   & $-6.33$ \\
USco\,1610-2239   & 2014-07-01 03:26UT & $\leq-5.92$ & $\leq-5.04$ & $\leq-6.20$ & $\leq-6.54$ & $-6.93\pm0.15$ & $-6.93\pm0.15$  & $-6.33$ \\
USco\,1610-2239   & 2014-07-01 04:04UT & $\leq-6.30$ & $\leq-5.99$ & $\leq-6.38$ & $\leq-6.77$ & $-6.96\pm0.07$ & $-6.96\pm0.07$   & $ -6.33$ \\
USco\,1608-2315 & 2014-06-15 & $\leq-5.98$ & $\leq-5.45$ & $-5.25\pm0.21$ & $-5.29\pm0.09$ & $-5.16\pm0.03$ & $-5.20 \pm 0.11$  & $-6.30$ \\
USco\,1608-2315 & 2014-06-20 & $\leq-5.98$ & $\leq-5.52$ & $-5.37\pm0.18$ & $-5.56\pm0.08$ & $-6.34\pm0.04$ &  $-5.99 \pm 0.10$  & $-6.30$ \\
USco\,1607-2211 & 2014-04-22	& $\leq-5.96$ & $\leq-5.92$ & $\leq-6.25$ & $\leq-6.52$ & $-6.57\pm0.07$ & $-6.57\pm0.07$   & $-6.27$ \\
USco\,1607-2211 & 2014-06-14	& $\leq-6.16$ & $\leq-6.06$ & $\leq-6.32$ & $\leq-6.52$ & \dots & \dots & \dots \\
\hline
\hline
\end{tabular}
\label{Tab:Lacc}
\tablefoot{The accretion luminosity for each line is derived from \cite{2012AeA...548A..56R}. The weighted mean of the $\mathrm{L_{acc}}$ values are reported in the "Balmer" column. The $\mathrm{log(L_{acc, noise}/L_{\odot})}$ value of USco\,1610-1913\,B assumes the companion is a single object. In case of higher multiplicity, and assuming a $\mathrm{L_{bol}}$ similar to HIP\,77900\,B, we find $\mathrm{log(L_{acc, noise}/L_{\odot})\,=\,-6.11^{+0.19}_{-0.22}}$.}
\end{table*}

\begin{table*}
 \centering
  \caption{Accretion rates from apparent fluxes of the emission lines for the companions and free-floating brown dwarfs from our original sample.}
\renewcommand{\arraystretch}{1.3}
\tiny
  \begin{tabular}{cc|ccccc|c}
  \hline   
	\hline
Object			& Date	& 	Ca II-H	&	Ca II-K	&	\Hg	&	\Hb	&	\Ha	&	10\%~\Ha 	  \\
				&       &  log($\Msun$/year)	& log($\Msun$/year)   & log($\Msun$/year) & log($\Msun$/year) &	log($\Msun$/year)	 &  log($\Msun$/year)	\\
\hline
USco\,1610-1913\,B & 2014-04-02    & $-12.10\pm0.53$  &  $-12.09\pm0.69$  &  $-12.53\pm1.02$  &  $-12.42\pm0.46$  &  $-12.14\pm0.37$ & $-11.48\pm0.40$  \\
HIP\,77900\,B	& 2014-04-12 05:45UT       & $<-11.69$  &  $<-11.82$  &  $<-11.84$  &  $<-12.23$  &  $-13.22\pm0.30$ & \dots  \\
HIP\,77900\,B	& 2014-04-12 08:38UT       & $<-11.90$  &  $<-11.80$  &  $<-12.15$  &  $<-12.22$  &  $-13.04\pm0.27$ & \dots  \\
HIP\,78530\,B	& 2014-06-09 04:08UT       & $<-11.20$  &  $<-11.11$  &  $<-11.38$  &  $<-11.34$  &  $-13.01\pm0.41$ & \dots  \\
HIP\,78530\,B	& 2014-06-09 05:14UT      & $<-11.52$  &  $<-11.25$  &  $<-11.61$  &  $<-11.42$  &  $-13.06\pm0.36$ & \dots  \\
\hline
USco\,1607-2242 & all epochs            & $<-12.32$  &  $<-11.99$  &  $<-12.62$  &  $<-12.85$  &  $<-13.57$  & \dots  \\
USco\,1608-2232 & all epochs	           & $<-12.56$  &  $<-11.45$  &  $<-13.08$  &  $<-12.85$  &  $<-13.45$ & \dots  \\
USco\,1606-2335 & 2014-06-25          & $<-11.86$  &  $<-11.83$  &  $-12.39\pm1.51$  &  $-12.52\pm0.80$  &  $-12.88\pm0.33$ & \dots  \\
USco\,1606-2335  & 2014-06-30         & $<-12.09$  &  $<-11.43$  &  $-12.20\pm1.18$  &  $-12.64\pm0.61$  &  $-12.73\pm0.26$ & $-11.25\pm0.42$  \\
USco\,1606-2335 & 2014-07-04          $<-12.21$  &  $<-12.00$  &  $-12.16\pm0.57$  &  $-12.25\pm0.33$  &  $-12.34\pm0.20$ & $-11.57\pm0.40$  \\
USco\,1606-2335 & 2014-07-29          & $<-12.15$  &  $<-12.00$  &  $-12.30\pm0.62$  &  $-12.54\pm0.37$  &  $-12.65\pm0.23$ & $-11.87\pm0.37$  \\
USco\,1610-2239   &  2014-06-30         & $<-12.29$  &  $<-12.20$  &  $<-12.47$  &  $<-12.89$  &  $-13.46\pm0.27$ & \dots  \\
USco\,1610-2239   & 2014-07-01 03:26UT  & $<-12.08$  &  $<-11.20$  &  $<-12.36$  &  $<-12.70$  &  $-13.27\pm0.33$ & \dots  \\
USco\,1610-2239   & 2014-07-01 04:04UT  & $<-12.46$  &  $<-12.15$  &  $<-12.54$  &  $<-12.93$  &  $-13.30\pm0.25$ & \dots  \\
USco\,1608-2315 & 2014-06-15            & $<-12.10$  &  $<-11.57$  &  $-11.59\pm0.43$  &  $-11.63\pm0.31$  &  $-11.50\pm0.25$ & $-10.76\pm0.45$  \\
USco\,1608-2315 & 2014-06-20            & $<-12.10$  &  $<-11.64$  &  $-11.71\pm0.40$  &  $-11.90\pm0.30$  &  $-12.68\pm0.26$ & $-10.86\pm0.45$  \\
USco\,1607-2211 & 2014-04-22	         & $<-12.08$  &  $<-12.04$  &  $<-12.37$  &  $<-12.64$  &  $-12.91\pm0.29$ & $-11.34\pm0.41$  \\
USco\,1607-2211 & 2014-06-14	         & $<-12.28$  &  $<-12.18$  &  $<-12.44$  &  $<-12.64$ & \dots & \dots  \\
\hline
\hline
\end{tabular}
\label{Tab:accretion_rate}
\tablefoot{The accretion rate for each line is derived from \cite{2017AeA...600A..20A} and the one from the 10\% \Ha\, is derived from \cite{2004AeA...424..603N}}
\end{table*}

\section{Discussion}
\label{Sec:Discussion}
	\subsection{Revisiting the physical association of HIP\,77900\,B and USco\,1610-1913\,B with \textit{Gaia}}
	\label{subsec:physasso}
HIP\,77900\,B  and USco\,1610-1913\,B are bright enough and distant enough from their primary so that they have reported \textit{Gaia} parallaxes. This is to our knowledge the first case for which the physical association of young imaged BD companions with their primary stars can be investigated based on the individual 5-parameters astrometric solutions of the system components.

\subsubsection{The case of HIP\,77900\,B}
The \textit{Gaia}-DR2 parallax of HIP\,77900\,A confirms that the system would be extreme if bound ($22.3\,\!''\,=\,3375\,$au projected separation). Very wide systems such as HIP\,77900 are expected to be rare \citep[e.g., ][]{2018arXiv180708799B}. The companion has not been observed at multiple epochs so it could in principle be a background star. However, the close resemblance of the companion's spectrum to those of free-floating analogues from our library of Upper-Sco objects identifies it as a likely member of this association. Therefore, HIP\,77900\, A and B are either coeval or aligned by chance within the association. Their respective distances ($191^{-30}_{+43}$\,pc for A, and $151.4_{-2.7}^{+2.8}$\,pc for B) reveal a 1.4\,$\sigma$ difference. There is in addition a 1.8\,$\sigma$ difference between the proper motion in declination of the two objects. The p-value on the difference on the 3 astrometric parameters of A and B (0.06) does not allow us to firmly conclude whether the system is coeval. The RUWE values of A and B (0.97 and 1.21) indicate robust solutions. The galactic Cartesian coordinates of HIP~77900\,A ($\mathrm{X_{A}\,=\,137.8\,\pm\,2.6}$\,pc, $\mathrm{Y_{A}\,=\,-35.5\,\pm\,0.7}$\,pc,  $\mathrm{Z_{A}\,=\,51.8\,\pm\,1.0}$\,pc) computed following \cite{2014ApJ...783..121G} are well within the range of expected values for Upper-Sco \citep{2018MNRAS.477L..50G},  contrary to the ones of B ($\mathrm{X_{B}\,=\,173.8^{+39.1}_{-27.3}}$\,pc, $\mathrm{Y_{B}\,=\,-44.7_{-10.1}^{+7.1}}$\,pc,  $\mathrm{Z_{B}\,=\,65.4^{+14.8}_{-10.3}}$\,pc) which is at odds with the youth of the object. However, we find photometric distances of $151^{+35}_{-24}$\,pc and  $150^{-17}_{+22}\,$pc for HIP\,77900\,B using the flux-calibrated spectra  of the spectral analogue M8.5 dwarfs from our sample (USco\,1607-2211 and USco\,1610-2239, respectively) with \textit{Gaia} parallaxes. These photometric distances are more consistent with the \textit{Gaia} distance of A and would strengthen the case of a coeval system. However, we found a RV of $19.3\pm1.2$\,km/s for the B component not consistent with the $1.3\pm2.6$\,km/s found by \cite{2006AstL...32..759G} for HIP\,77900\,A. This result raises severe doubts about the gravitational link between both objects, and the next \textit{Gaia} data releases of the relative astrometry of A and B will be needed to firmly conclude on the physical association of both objects.

\subsubsection{The case of USco\,1610-1913\,B}
The spectral properties and measured distance of USco\,1610-1913\,B clearly confirm its membership to Upper-Sco. The two objects have been shown to have a common proper motion from 2007 to 2012 and have been proposed to be bound \citep[][and ref therein]{2014ApJ...781...20K}. However, the recent DR2 solutions of  USco\,1610-1913\,B and A diverge in proper motion (2.0 and 2.4\,$\sigma$ significance in $\mathrm{\mu_\alpha}$ and $\mathrm{\mu_\delta.cos(\delta)}$, respectively) and in distance (1.4\,$\sigma$; B is 10.3\,pc closer than A). A $\chi^{2}$ test on the difference on the 3 astrometric parameters of each system component (parallaxes and proper motions) taking into account the correlations, gives a p-value of 0.01 which is in favor of a physical bond between both objects. The p-value accounts for the slight under-estimation of the errors  \citep{2018AeA...616A...2L}. The reliability of the \textit{Gaia} astrometry may however be questioned for this particular system. Indeed, the  $\sim0.145\,\!''$ M-dwarf  companion to USco\,1610-1913\,A \citep{2008ApJ...679..762K} is unresolved in the DR2. The Re-normalised Unit Weight Error (RUWE) index is proposed as a reliable and informative goodness-of-fit statistic than for instance the astrometric excess noise \citep[Gaia technical note Gaia-C3-TN-LU-LL-124-01]{2018AeA...616A...2L}. The RUWE of USco\,1610-1913\,A (1.63) is above the threshold of 1.4 and indicates that the observations are inconsistent with a simple 5-parameter astrometric model while B (RUWE=1.19) shows a more reliable solution. The next releases of \textit{Gaia} will here again solve this ambiguity.

If we keep the hypothesis that the system is bound (or even coeval) and located at the distance of the B component, USco\,1610-1913\,B remains 4.5 times more luminous than free-floating analogues from the Upper Sco association. This is illustrated in Figure \ref{Fig:Overlum} using for comparison USco\,1607-2211 and USco\,1610-2239 which both have high quality astrometric solutions. In addition, the probability for a chance-alignment within the association is slim. \cite{2013ApJ...773...63A} identified USco\,1610-1913\,B while searching for distant companions within 30 arcseconds from Upper Sco stars and with optical (Pan-Starrs) and near-infrared (UKIDSS) colors compatible with young, cool objects. Using the now available Pan-Starrs data \citep{2016arXiv161205560C}, we confirm that USco\,1610-1913\,B belongs to one of the three objects with the reddest $i-z$ and $i-y$ colors  within 5 arcmins of  USco\,1610-1913\,A. It is also the only object in this field with colors typical of M7-M9  Upper-Sco objects. \cite{2012ApJ...745...56D} determined a RV of $−6.91\pm0.27$\,km/s for USco\,1610-1913\,A. USco\,1610-1913\,B is the unique target for which with find a strong discrepancy between the RV determined by \texttt{ForMoSA} and the cross-correlation approach ($11.5\pm0.4$\,km/s and 0.9 km/s, respectively). Both values are in addition inconsistent with the RV derived by \cite{2012ApJ...745...56D} for the A component, and therefore inconclusive. Although we cannot firmly exclude that USco 1610-1913 A and B are simply chance aligned members of Upper-Sco, parallaxes and common proper motion do support that they form at least a comoving and possible coeval pair.

\subsection{The over-luminosity of USco\,1610-1913\,B}
\label{subsec:overlum}
Evolutionary models of \cite{2000ApJ...542L.119C} predict a \Teff\,=\,2827$\,\pm\,$169 K for USco\,1610-1913\,B using the bolometric luminosity of the source. This corresponds to a M5 brown dwarf (see Table \ref{Tab:param_compare}), incompatible with the spectral type of M9$\,\pm\,$0.5 derived from our empirical analysis. Our results using the \texttt{ForMoSA} code (see Section \ref{Sec:Results}) further confirms this over-luminosity, based on model atmosphere fits.  There are several other sources in literature which are similarly over-luminous, such as USco\,1602−2401\,B \citep{2013ApJ...773...63A} or 2MASS\,J162243.85−195105.7 \citep{2012ApJ...745...56D}. \cite{2013ApJ...773...63A} proposed that differences in the accretion history of USco\,1610-1913\,B could play a role in such a discrepancy as proposed by \cite{2012ApJ...756..118B}. For late K and M-dwarfs, chromospheric activity as found in our spectra decreases the objects \Teff\,and increases their radii \citep{2007ApJ...660..732L, 2008AeA...478..507M}. Based on a sample of 669 M$\,<1\,\Msun$ non-accreting stars from the Palomar/Michigan State University catalog (PMSU; \citealt{1995AJ....110.1838R}; \citealt{1996AJ....112.2799H}), \cite{2012ApJ...756...47S} proposed empirical relations to determine the bias induced by chromospheric activity :

\begin{equation}
\frac{\Delta~T_{eff}}{T_{eff}}~(\%)~=~(-3.12~\pm~3.15)~\times~(log\left(\frac{L_{H_{\alpha}}}{L_{bol}}\right)~+~4)~+~(-5.1~\pm~0.7)
\end{equation}

\begin{equation}
\frac{\Delta~R}{R}~(\%)~=~(8.00~\pm~7.63)~\times~(log\left(\frac{L_{H_{\alpha}}}{L_{bol}}\right)~+~4)~+~(11.2~\pm~1.6)
\end{equation}

Despite this variations of the atmosphere's radius and \Teff, the bolometric luminosity remains constant. This effect can't explain the over-luminosity of USco\,1610-1913\,B. An alternative explanation would be that the companion is an unresolved quadruple system presumably with all components having similar spectral types, in this case. Such systems are expected to be rare among A and FG-type stars \citep{2010ApJS..190....1R, 2014MNRAS.437.1216D, 2014AJ....147...87T}. The system would be even more exotic and rare, considering that USco\,1610-1913\,A is a binary and that USco\,1610-1913\,B has to be composed by four identical objects to account for its overluminosity.
Quadruples tend to be found as 2+2 tight binary systems. The discovery of a quadruple system made of two pairs of M5 eclipsing binaries in Upper-Sco \citep{2018ApJ...865..141W} show that such object exist in isolation. To our knowledge, adaptive-optics images of the system have not displayed PSF-elongation nor resolved the companion as a higher-order object but new near-infrared high-resolution (R~80 000 to 100 000) spectrographs (ESO/NIRPS, CRIRES+) could investigate the multiplicity of USco\,1610-1913B in the near-future.

\section{Conclusion}
Based on medium resolution spectra obtained with the X-Shooter spectrograph at VLT, we carried out an in-depth characterization of three low-mass brown dwarf companions on wide-orbits, specifically USco\,161031.9-16191305\,B, HIP\,77900\,B, and HIP\,78530\,B of the Upper-Scorpius association, together with six young isolated brown dwarfs of similar spectral types and ages.  The X-Shooter data yield the first medium-resolution optical spectra of the companion objects. We can summarize the main results as follow:
\begin{enumerate}

\item the re-investigation of the spectral classification of the three companions USco\,1610-1913\,B, HIP\,77900\,B and HIP\,78530\,B confirms that they have spectral types M$9\,\pm\,0.5$, M$9\,\pm\,0.5$, and M8$\,\pm\,0.5$, respectively. HIP\,77900\,B and HIP\,78530\,B are identified as young, intermediate surface gravity brown dwarfs, whereas USco\,1610-1913\,B is confirmed as a very-low surface  gravity  brown dwarf.

\item the development and use of the \texttt{ForMoSA} forward modelling code relying on the Nested Sampling procedure enables us to infer posterior probability distributions of the physical properties (\Teff, log(g), $R$, $L$ and extinction) of USco\,1610-1913\,B, HIP\,77900\,B and HIP\,78530\,B using the \texttt{BT-SETTL15}\ atmospheric models and the X-Shooter spectra. We find that generally, the models fail to reproduce the pseudo-continuum of the X-Shooter  spectra over broad wavelengths range. Our solutions are mainly affected by the choice of the spectral range considered to estimate the best fit, and the fitting error bars remain about one to two orders of magnitude smaller given the high S/N of the X-Shooter spectra. Finally, when the extinction  is considered as a free parameter, the \texttt{ForMoSA} fitting solutions are considerably improved at all wavelengths with extinction values of 1.6-2.6\,mag suggesting a clear deficiency in the dust grain modeling of the \texttt{BT-SETTL15} atmospheric models. These inconsistencies limit our ability to investigate at this stage the chemical abundance of heavy elements in  the  atmosphere of these young brown dwarfs that could trace different formation mechanisms. Finally, the key physical properties such as \Teff, log(g), $L$ are in agreement with the empirical analysis, but indicate a clear over-luminosity for USco\,1610-1913\,B.

\item the study of the medium resolution optical part of the X-Shooter spectra allowed us to identify the presence of various Balmer lines  for the three companions USco\,1610-1913\,B, HIP\,77900\,B and HIP\,78530\,B, and the two isolated brown dwarfs USco\,1608-2315 and USco\,1607-2211, and to investigate their origin. Their low accretion rate, low accretion luminosity and low 10\% width tend to support them as signatures tracing chromospheric activity except for USco\,1608-2315, which adds to the limited population of accreting free-floating young brown dwarf with mass below 30\,\MJup. 

\item the nature of USco 1610-1913 B and HIP 77900 B is then revisited and discussed in the context of the new \textit{Gaia} DR2 results. The \textit{Gaia} solutions of USco 1610-1913\,A are probably affected by the binarity of USco 1610-1913\,A itself. For  HIP 77900\,B, the parallax is surprisingly inconsistent with the photometric distance. Nevertheless, a coeval (bound or unbound) configuration remains the most plausible one for both systems. Finally, we showed that the over-luminosity of USco\,1610-1913\,B cannot be explained by chromospheric activity suggesting that it might be a high-order multiple component if we want to reconcile spectral type and observed luminosity. It calls for further studies using high-resolution spectrographs ($R\mathrm{_{\lambda} \simeq 80 000 - 100 000}$) like ESO3.6m/NIRPS or VLT/CRIRES+ to explore such an hypothesis.

\end{enumerate}

\begin{acknowledgements}
The authors thank the ESO staff for conducting the observations. We are grateful to D. Lachapelle for providing the spectra of HIP\,78530\,B and USco\,1610-1913\,B. We also thank J. Bouvier for advising us on accretion process and chromospheric activity effects. We acknowledge financial support from the Programme National de Plan\'{e}tologie (PNP), the Programme National de Physique Stellaire (PNPS) of CNRS-INSU in France, and the Agence Nationale de la Recherche (GIPSE project; grant ANR-14-CE33-0018). This research has made use of the SIMBAD database and VizieR catalogue access tool (operated at CDS, Strasbourg, France). This research has made use of NASA's Astrophysics Data System and of the Extrasolar Planet Encyclopedia (http://exoplanet.eu/). This publication makes use of VOSA, developed under the Spanish Virtual Observatory project supported from the Spanish MINECO through grant AyA2017-84089. G-DM acknowledges the support of the DFG priority program SPP 1992 ``Exploring the Diversity of Extrasolar Planets'' (KU 2849/7-1).
\end{acknowledgements}

\begin{appendix}
\section{Temperature and reddenning of USco\,1610-1913.}
\label{sec:AppA}
We built the spectral energy distribution (SED) of USco\,1610-1913 from SDSS \citep[u, g, r, i bands][]{2015ApJS..219...12A}, APASS \citep[B and V bands][]{2016yCat.2336....0H}, Pan-Starrs \citep[y band][]{2016arXiv161205560C}, 2MASS \citep[J, H, Ks][]{2003tmc..book.....C} and WISE \citep[W1, W2, W3 bands][]{2013yCat.2328....0C} photometry collected through the VOSA \citep{2008AeA...492..277B} web interface\footnote{http://svo2.cab.inta-csic.es/theory/vosa/index.php} supplemented by  Spitzer (8\mic, 24\mic) photometry  taken from \cite{2006ApJ...651L..49C}. 

We compared deredenned SED of the object to synthetic spectra from the BT-NEXTGEN library \citep{2012RSPTA.370.2765A} with \Teff\,in the range 3500-4000K and M/H\,=\,0. The surface gravity was varied from 3.0 to 5.0 dex and does not influence the fit. We considered $\mathrm{A_{v}}$ in the range 0-5 mag in steps of 0.01 mag and a reddenning law \citep{2003ARAeA..41..241D, 2003ApJ...598.1026D, 2003ApJ...598.1017D}  with a reddenning parameter $\mathrm{R_{v}\,=\,A_{v}/E(B-V)\,=\,3.1}$. 

\begin{figure}
  \centering
    \includegraphics[width=\columnwidth]{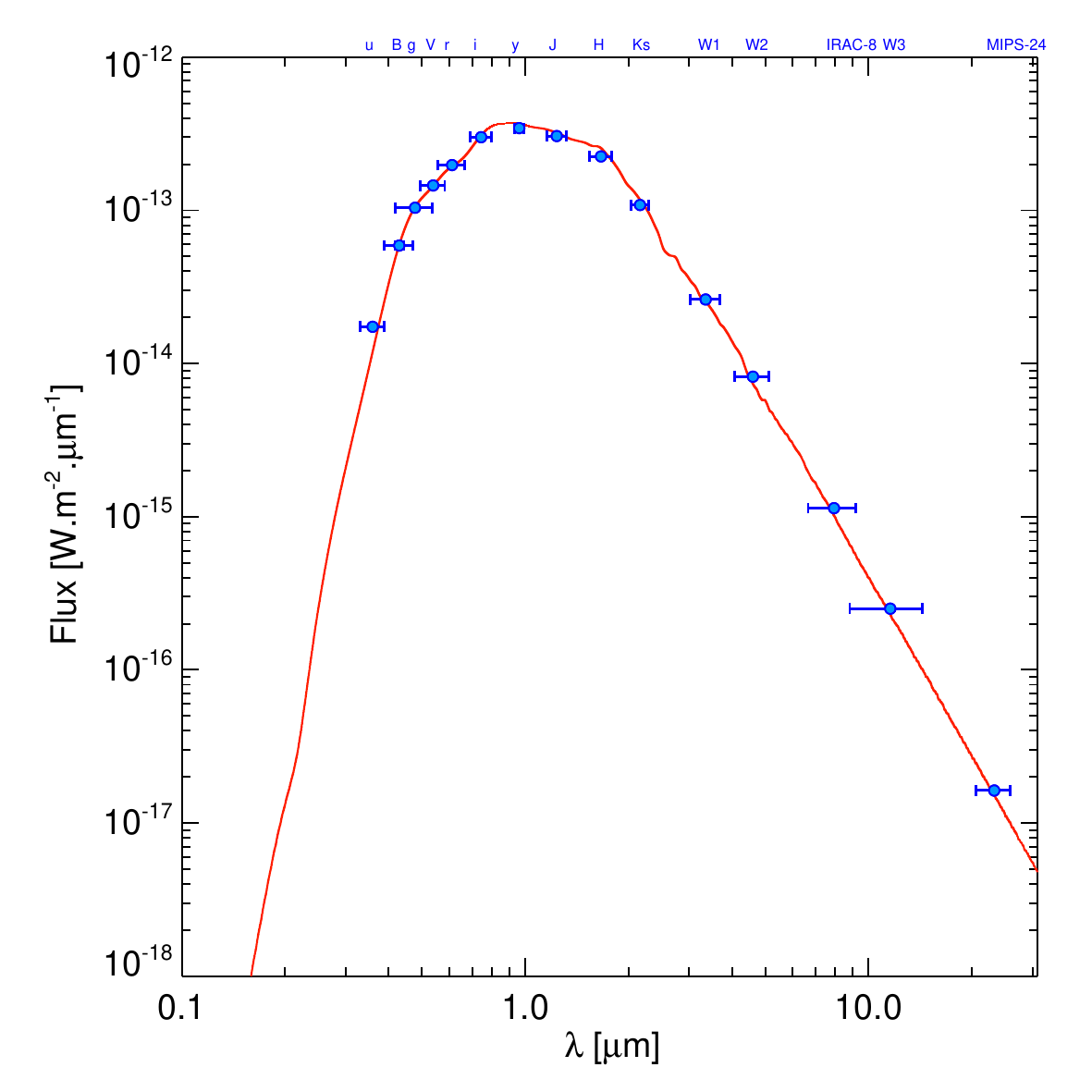}
  \label{Fig:SEDstar}
	\caption{Spectral energy distribution of USco\,1610-1913 dereddenned by $\mathrm{A_{v}\,=\,0.09}$ mag compared to the best-fitting BT-NEXTGEN model (\textit{red} line) corresponding to \Teff\,=\,3800K, log g\,=\,4.0 dex, and M/H\,=\,0 dex.}
\end{figure}

We find a best fit for \Teff\,=\,3800$\,\pm\,$100 K and $\mathrm{A_{v}}$\,=\,0.09$\,\pm\,$0.01 mag) (Fig. \ref{Fig:SEDstar}). The temperature is in broad agreement \citep{2013ApJS..208....9P} with that expected for the quoted K7 spectral type \citep{2001AJ....121.1040P}  and rather suggests a M0 type for the object. We confirm the lack of an excess up to 24 \mic. The fit clearly excludes solutions at a higher \Teff\,and a reddenning of $\mathrm{A_{v}}$\,=\,1.1 mag as reported by \cite{2009ApJ...705.1646C}.

\section{Systematic differences in spectral type determination.}
\label{sec:AppB}
\begin{figure}
  \centering
    \includegraphics[width=\columnwidth]{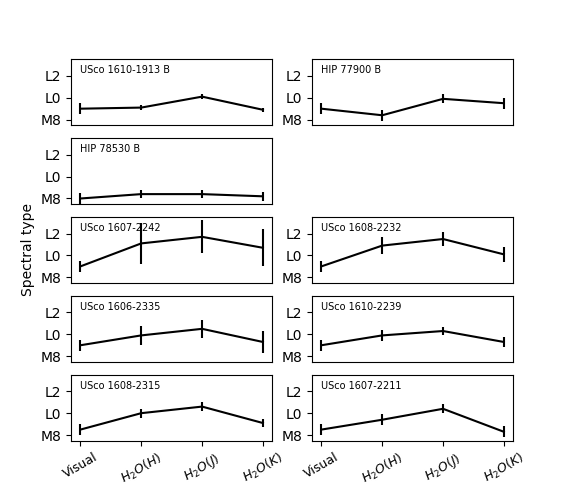}
  \label{Fig:plot_spec_type}
	\caption{Determination of the spectral type in using $\mathrm{\chi^{2}}$\, comparison in optic and H$_{2}$O absorption bands.}
\end{figure}

To determine the spectral type of our sources we focused on the visible in using a standard $\mathrm{\chi^{2}}$\, comparison between our spectra and the Ultracool RIZzo Spectral Library (see Section \ref{Sec:Empirical_analysis}). We also used relations between the spectral type and the H$_{2}$O absorption bands from \cite{2007ApJ...657..511A} (H$_{2}$O (H)) and \cite{2004ApJ...610.1045S} (H$_{2}$O (J) ; H$_{2}$O {K}). Figure \ref{Fig:plot_spec_type} shows the systematic differences according to each method used to determine the spectral type.

\end{appendix}

\bibliographystyle{aa}
\bibliography{ms}

\end{document}